\documentclass[aps,twocolumn, prd,
    superscriptaddress,
    showkeys,showpacs,
    nofootinbib,
   nobibnotes]{revtex4-2}
\usepackage{natbib}
\usepackage[pdftex]{graphicx}
\usepackage[dvipsnames]{xcolor}
\usepackage{xcolor}
\usepackage{amsmath}
\usepackage{amssymb}
\usepackage{dcolumn}
\usepackage{bbold}
\usepackage{braket}
\usepackage{lipsum} 
\usepackage{caption}
\usepackage{subcaption}
\usepackage{amsthm}
\usepackage{setspace}
\usepackage{hyperref}
\usepackage{soul}
\usepackage{graphicx}
\usepackage[T1]{fontenc}
\usepackage{enumitem}
\definecolor{darkblue}{RGB}{0,0,149}
\hypersetup{
	colorlinks=true,
	linkcolor=red,
	filecolor=darkblue,
	urlcolor=darkblue,
	citecolor=Green,
	linktocpage=true,
}
\usepackage{url}

\newcommand{\QCF}[1]{X-QRF}

\begin{document}
\title{Quantum Reference Fields Transformations in Linearized Quantum Gravity}

	\author{Lin-Qing Chen}%
	\email{linqing.nehc@gmail.com}

    \affiliation{Institute for Quantum Optics and Quantum Information (IQOQI),Austrian Academy of Sciences}
    \affiliation{University of Vienna, Faculty of Physics, Vienna Center for Quantum Science and
Technology, Boltzmanngasse 5, 1090 Vienna, Austria}
	\affiliation{Institute for Theoretical Physics, ETH Z{\"u}rich, 8093 Z{\"u}rich, Switzerland}
	
	\author{Flaminia Giacomini}%
	\email{flaminia.giacomini@uniroma2.it}
	\affiliation{Institute for Theoretical Physics, ETH Z{\"u}rich, 8093 Z{\"u}rich, Switzerland}
\affiliation{Dipartimento di Fisica, Universit{\`a} di Roma Tor Vergata and Sezione INFN Roma2, Via della Ricerca Scientifica 1, 00133, Roma, Italy}

    \begin{abstract}  
    Diffeomorphism invariance is a central feature of general relativity.    Without external reference structures, matter and geometry must be specified relationally, with respect to internal subsystems serving as reference frames. In quantum gravity, these reference systems must themselves be treated as quantum, motivating the use of quantum reference frames.
    In this work, we address how such a relational description could be formulated within linearized quantum gravity. To this purpose, we introduce quantum reference fields, i.e.\,sets of four dynamical scalar fields whose stress-energy tensors enter the gravitational constraints. 
    These fields extend the notion of quantum reference frames to local field-theoretic reference systems, allowing matter and gravitational degrees of freedom to be described relationally with respect to physical quantum systems whose states cannot be  sharply peaked on definite field configurations. By generalizing the perspective-neutral construction of quantum reference frames, we show that relational, gauge invariant observables admit reduced descriptions in the perspective of each quantum reference field, and we derive the unitary transformations relating them. 
    The resulting unitary maps implement local quantum coordinate changes between different internal perspectives, and act on the linearized gravitational field with an analogous structure to a linearized diffeomorphism, but with the classical gauge parameter replaced by a physical quantum field. 
    Finally, we construct a relational von Neumann-type measurement scheme,  showing how the corresponding reduced observables can be accessed operationally from the perspective of a quantum reference field. 
	\end{abstract}

	\maketitle

\section{Introduction}

Diffeomorphism invariance lies at the core of general relativity and shapes the very notions of physical observables~\cite{Rovelli1991,Rovelli:2001bz, Dittrich:2005kc, Dittrich2007,Tambornino:2011vg, Brunetti:2013maa, Giddings:2025xym, Baldazzi:2021fye,Goeller:2022rsx}, locality~\cite{Donnelly:2016rvo,Goeller:2022rsx}, time~\cite{dewitt1967quantum, Isham:1992ms,Anderson:2010xm} and symmetries~\cite{1962RSPSA.269...21B,PhysRev.128.2851, PhysRevLett.10.66, 1986CMaPh.104..207B, Donnelly:2016auv,Strominger2017}. Its realisation in the quantum regime is a central conceptual and technical challenge for quantum gravity, arising in different ways across existing frameworks~\cite{Ashtekar:1991hf,Rovelli:2004tv,Thiemann2007,Oriti2009, PhysRevD.83.104051, FREIDEL2003279, Donnelly:2016rvo,Chen:2016aag, Dowker:2019qiz, Steinhaus:2020lgb,Witten:2023xze}. Since quantum spacetime is unlikely to admit a fundamental description in terms of a differentiable manifold, we need to re-examine the notion of diffeomorphisms in the quantum regime.

It has long been suggested that a background-independent description of quantum spacetime should be relational~\cite{Rovelli:1990pi}. In the absence of an external reference frame, one is naturally led to an internal description relative to physical subsystems, with matter degrees of freedom  serving as the reference~\cite{Rovelli:1990pi, Kuchar:1990vy,Brown:1994py,Brown:1995fj,Giesel:2007wn, Giesel:2012rb,Giesel:2016gxq, Hoehn:2023axh,Thiemann:2026vaj}. 
Due to the universal coupling between gravity and matter, the back-reaction of these reference systems is fundamentally non-negligible once gravity itself is quantum, and must therefore be incorporated consistently into the gauge structure inherited from classical general relativity. 
However, a coherent formulation of quantum reference transformations associated with physical fields, capable of enacting local transformations on quantum gravitational states, remains an outstanding open question.

In this work, we show that the above question can be addressed in linearized quantum gravity, providing the simplest setting in which the gravitational field can be treated quantum mechanically, and  the weak-field, low-energy regime that any viable high-energy theory of quantum gravity is expected to reproduce. Our construction builds on the framework of quantum reference frames (QRFs), 
which has been developed in parallel since the 1960s in quantum gravity~\cite{dewitt1967quantum} 
and in quantum information (QI) and quantum foundations (QF)~\cite{aharonov1, aharonov3}. The discussion has recently been revived, showing convergence between different research directions~\cite{brs_review, spekkens_resource, palmer_changing, katz2015mesoscopic, smith_quantumrf, poulin_dynamics, busch_relational_1, busch_relational_2, busch_relational_3, angelo_1,Giacomini:2017zju, Vanrietvelde:2018pgb, perspective2,  hoehn2019trinity,hoehn2019switching, giacomini2019relativistic, hoehn2020equivalence,  ballesteros2020group, streiter2020relativistic,  yang2020switching, castro2020quantum, de2020quantum, krumm2020, tuziemski2020decoherence, mikusch2021transformation, hoehn2021quantum, hoehn2021internal, castro2021relative, delaHamette:2021oex, de2021entanglement, giacomini2021spacetime, Carette:2023wpz, Glowacki:2023nnf, Hoehn:2023ehz}.

The recent development of QRFs at the interface of QF and gravity has yielded  conceptual insights into quantum spacetime. For instance, the QRF transformation developed for the superposition of effectively classical geometries~\cite{Hardy:2018kbp, giacomini2020einstein, giacomini2021quantum,delaHamette:2022cka, de2021falling, Kabel:2022cje, Kabel:2024lzr} has been used to formulate an extension of Einstein's Equivalence Principle \cite{hardy2020implementation, giacomini2020einstein, giacomini2021quantum, cepollaro2021quantum}, quantum diffeomorphisms ~\cite{Hardy:2018kbp, hardy2020implementation, giacomini2020einstein, giacomini2021quantum,delaHamette:2022cka, Kabel:2024lzr}, the notion of events and relative localisation \cite{castro2020quantum, Kabel:2024lzr, Vilasini:2025qun}. In a broader context,  it was shown that reference frames on spacetime boundaries can encode large gauge transformations 
~\cite{Donnelly:2016auv, Strominger2017, Carrozza:2022xut, Kabel:2023jve, Janssen:2025uzf, fewster2025semilocalobservablesedgemodes}; 
the inclusion of 
time reference frames is essential in 
the type reduction of von Neumann algebras and defining entropy in quantum field theory and gravity~\cite{Witten:2022CrossedProduct, Chandrasekaran:2023DeSitterAlgebra, Fewster:2024pur, DeVuyst:2025CrossedProductsQRFs, DeVuyst:2024khu}. Finally, constructing QRFs out of the gravitational field itself has enabled a fully covariant quantisation of null rays~\cite{Wieland:2021vef, Freidel:2025ous, Wieland:2025qgx, Freidel:2026stu}. All these developments point to the necessity of extending the logic of QRFs to perform local transformations acting jointly on matter and gravity degrees of freedom.
 
In this work, we generalize QRFs by associating them to physical quantum fields which are included in the gravitational constraints. These fields serve as quantum coordinate systems, with respect to which we derive an internal, relational description of other physical fields. We identify the corresponding relational observables—including those of the linearized quantum gravitational field—and construct a unitary transformation that consistently relates different perspectives. We further constructed a relational measurement scheme for the relational observables in the reduced perspective. 
This work provides a necessary and crucial step towards the long-term goal of achieving a background-independent formulation of quantum spacetime.

\section{quantum reference fields in linearized gravity}

A set of quantum reference fields (\QCF{}s)  corresponds to a set of four scalar quantum fields $\hat{X}^{(\mu)} $\footnote{The notation $(\mu)$ labels different scalar fields which represent
different spacetime dimensions.}  with $\mu=0,1,2,3$, defined on an open subregion $U \subset \mathcal{M}$ of a spacetime manifold $\mathcal{M}$. We further use  an auxiliary coordinate chart  $s: U  \rightarrow \mathbb{R}^4$  to parametrize these physical  fields.
\begin{figure}[h]
\includegraphics[width=0.45\textwidth]{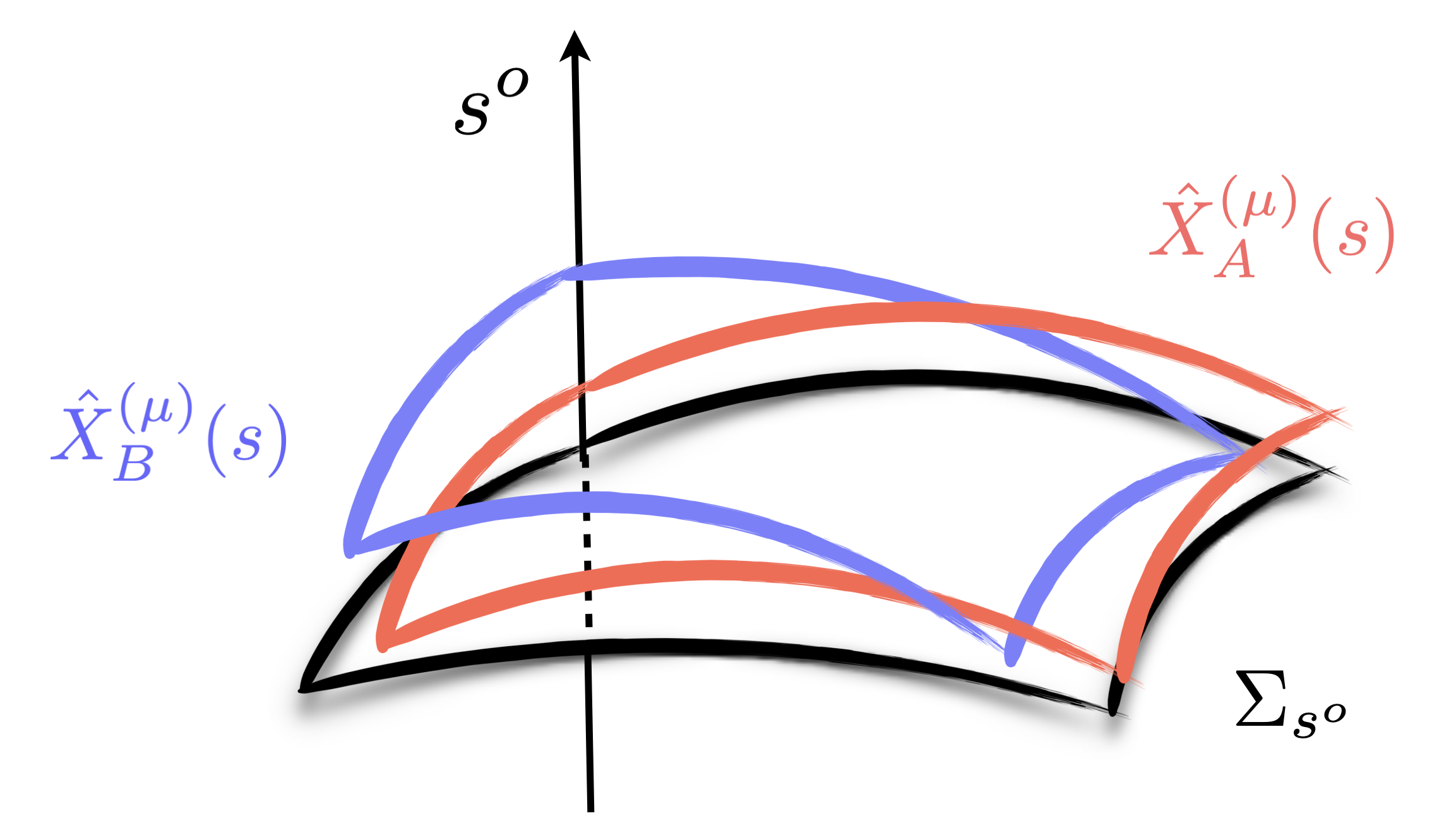}
    \caption{Two quantum coordinate fields parametrized by an auxiliary coordinate $s$.}
\end{figure}
To connect the spectrum of $\hat{X}^{(\mu)}$ to  the familiar notion of coordinates, it is convenient to express the quantum state of \QCF{}s  in the field basis~\cite{Hatfield:1992rz}.  For canonical analysis, we focus  on an arbitrary equal-time hypersurface $\Sigma_{s^0}$ defined with respect to the auxiliary coordinates $s = (s^0, \vec{s})$. Formally, $\hat{X}^{(\mu)}$ are operator-valued distributions, smeared on an equal-time  slice using normalized smooth functions $f_{\sigma,\mathrm{s}}$. Such functions are centred at $\mathrm{s}$ and have characteristic spatial width $\sigma$, which sets the spatial resolution. 
$\hat{X}^{(\mu)}_{f_s} (s^0) := \int d^3\vec{u} f_{\sigma,\mathrm{s}}(\vec{u}) \hat{X}^{(\mu)} (s^0, \vec{u})$.
The eigenbases $\{|x^{(\mu)} \rangle \}_{\mu = 0, 1,2,3}$ of the field operators $\hat{X}_{f_s}^{(\mu)}(s)$ are defined as
 	\begin{equation} \label{coordeigen}
 		\hat{X}_{f_s}^{(\mu)} (s^0)  |x^{(\mu)} \rangle = x_{f_s}^{(\mu)} (s^0, \vec{s})  |x^{(\mu)} \rangle, 
 	\end{equation}
 where $\ket{x^{(\mu)}}$ are improper eigenstates normalized to a  delta functional, analogous to the improper eigenvectors of the position and momentum operator in standard quantum mechanics. These eigenfunctions are not necessarily smooth and invertible functions a priori. For notation brevity,  we denote $\hat{X}^{(\mu)}_{f_s}(s^0)$ by $\hat{X}^{(\mu)}(s)$ in the remainder of the paper whenever the smearing function need not be specified and no ambiguity arises.

In general, the quantum state of the reference fields can be expanded in the field basis as $ |\psi \rangle = \int D[x] \psi (x)  | x\rangle$, with $x:= x^{(\mu)}, \mu=0,1,2,3$ for brevity, 	and $D[x]$ is the covariant functional integration measure on  $\Sigma_{s^0}$.
In particular, $\hat{X}^{(0)}$ plays the role of a physical clock field. For a generic \QCF{} state, $\hat{X}^{(0)}$ may be in a superposition of clock readings on $\Sigma_{s^0}$.  In a later section, we derive the  conditions that quantum states of the X-QRFs must satisfy.

We now couple the \QCF{}s to linearized gravity around the Minkowski background, with $g_{\mu\nu}=\eta_{\mu\nu}+ h_{\mu\nu}, |h_{\mu\nu}|\ll 1$. Concretely, the total system we consider consists of a quantum matter source $S$, the linearized gravitational field $\hat{h}_{\mu\nu}$, and two sets of \QCF{}s $\hat{X}^{(\mu)}_A$ and $\hat{X}^{(\mu)}_B$.  We choose the X-QRF fields to have dimensions of length, matching the background coordinates. Expanding  the Arnowitt-Deser-Misner (ADM) Hamiltonian \cite{PhysRev.116.1322} around this background  and keeping terms up to quadratic order,
the total Hamiltonian  takes the form~\cite{Barnich:2008ts,Anastopoulos:2013zya, Chen:2022wro} 
\begin{equation} \label{eq:FullHam}
\begin{split}
&\hat{H}_{tot} =    \hat{H}_M +  \hat{H}_G -\frac{1}{2}\int d^3s \hat{h}_{ij} (\vec{s})\hat{T}_M^{ij} (\vec{s}) \\
    + &\int d^3s \,\left(\hat{n}(\vec{s})  \hat{\mathcal{C}} (\vec{s})+ \hat{n}_i (\vec{s}) \hat{\mathcal{C}}^i(\vec{s})  + \lambda_{\mu} (\vec{s}) \hat{\pi}^{0\mu} (\vec{s})  \right),
\end{split}
\end{equation}
where $\hat{H}_G$ is the free Hamiltonian of the linearized gravitational field, while $\hat{H}_M$ denotes the free Hamiltonian of all matter degrees of freedom.
The total matter stress-energy tensor $\hat{T}^M_{\mu\nu}$ contains the ordinary matter sector together with two \QCF{} sectors \, $\hat{T}^M_{\alpha\beta} = \hat{T}_{\alpha\beta}^S + \hat{T}_{\alpha\beta}^A+ \hat{T}_{\alpha\beta}^B$,  and $\hat{H}_M = \hat{H}_S+ \hat{H}_A +\hat{H}_B$. For  each set of \QCF{}s, the stress-energy tensor is the sum of the
contributions from its four scalar components, i.e. $\hat{T}_{\alpha\beta}^A = \sum_{\mu=0}^3 \hat{T}_{\alpha\beta}^{A(\mu)}$, $\hat{H}_{A} = \sum_{\mu=0}^3  \int d^3 s\, \hat{T}^{00}_{A(\mu)}$. For notational simplicity, we suppress the component labels and write $\hat{X}:= \hat{X}_A^{(\mu)}$; the contribution of a single scalar component is $T_{\alpha \beta}^{A}=   \Upsilon \partial_\alpha \hat{X} \partial_\beta \hat{X} - \eta_{\alpha \beta} \left(\frac{\Upsilon}{2}  \partial_\rho \hat{X} \partial^\rho \hat{X}  + V(\hat{X})\right)$, where $\Upsilon$ is a dimensionful free parameter and the potential $V(\hat X)$ is  included for generality. The B-sector is defined analogously.

The  four secondary first-class  constraints generate linearized  diffeomorphisms:
\begin{equation} \label{eq:constraintsLQG}
    \begin{split}
        &\hat{\mathcal{C}}(\vec{s}) : =  \partial_i \partial^i \hat{h}^T(\vec{s}) + \kappa \hat{T}_M^{00}(\vec{s}), \\
        &\hat{\mathcal{C}}^i(\vec{s}) : = 2 \partial_j \hat{\pi}_G^{ij} (\vec{s})+ \hat{T}_M^{0 i}(\vec{s}),
    \end{split}
\end{equation}
in which $\kappa = 16\pi G c^{-4}$, $h_T:= \delta^{ij} h_T^{ij}$ denotes the trace of  the transverse part of the spatial metric perturbation $ h_T^{ij} (\vec{k}):=P^i_k P^j_l h^{kl} (\vec{k}) $ and $P^i_j = \delta^i_j - \frac{k^i k_j}{|\vec k|^2}$ is the transverse projector.  
In addition, we work in the extended ADM phase space~\cite{Pons:1996av}, in which the perturbations of
lapse $n(\vec{s})$ and shifts $n_i(\vec{s})$ are retained as canonical variables. Their canonical momenta give the primary constraints $\pi^{0\mu} =0$, which  are formally imposed in the Hamiltonian Eq.(\ref{eq:FullHam}) by the Lagrangian multiplayer $\lambda_{\mu}$. (see Appendix~\ref{app:Enlarged} for details).  
This extension is equivalent to the original formulation  in terms of the physical solutions, but is needed for constructing the full generators of the diffeomorphisms transformations in the canonical picture. 
The  generators of linearized diffeomorphisms along an arbitrary vector field $l^{\mu}$ are
\begin{equation}\label{Main:fullgen}
\begin{split}
G_{l}(s^0) &= \int d^3s  \left( \pi_{0\mu}(\vec{s}) \dot{l}^{\mu}(s^0, \vec{s}) +\mathcal{C}_\mu(\vec{s}) l^{\mu} (s^0, \vec{s}) \right)\\+ & \int d^3x  d^3y \int d^3z  \tilde{K}^{\nu}_{\mu0} (x,y;z)  l^{\mu}(s^0, \vec{y}) \pi_{0\nu} (\vec{z}).
\end{split}
\end{equation}
in which $\tilde{K}^{\nu}_{\mu0}$ are the structure functions evaluated on the flat background.  The non-perturbative version was  derived in Ref.\,\cite{Pons:1996av}. 
The  diffeomorphisms generated by a vector field $l$ is given by $\delta_l g_{\mu \nu} (s)= \{G_{l}(s^0), h_{\mu \nu} (s^0,\vec{s})\} = \partial_{\mu} l_{\nu} (s)+ \partial_{\nu} l_{\mu}(s)$.

The dynamical constraints of the theory are satisfied if, on the physical Hilbert space, we have
\begin{equation} \label{eq:fullCgravQRF}
    \hat{\mathcal{C}}(\vec{s}) \ket{\Psi}_{ph} =0, \hspace{0.3cm} \hat{\mathcal{C}}^i (\vec{s}) \ket{\Psi}_{ph} =0.
\end{equation}
In addition, the extended phase-space formulation requires us to also impose $\hat{\pi}^{0\mu}(\vec{s})\ket{\Psi}_{ph} =0$. With static quantum  source, the Schrödinger representation of the physical state  was solved in \cite{Chen:2022wro, Chen:2024xvm}. In this work, to keep the time dependence explicit for the transformation of operators, we henceforth work in the Heisenberg picture. To keep the notation simple, we do not add a label but, unless explicitly stated, all quantities should be understood in the Heisenberg picture from now on.

\section{The admissible states of  quantum reference fields}
\label{sec:admissibleXQRF}

The perturbative regime of gravity restricts the admissible states of  \QCF{}s. 
We require that expressing the gravitational field with respect to the \QCF{}s must remain compatible with  the weak-field expansion.
Classically, this means that any set of \QCF{}s $ \hat{X}^{(\mu)}_A$  needs to satisfy the condition (see Appendix~\ref{app:CoordFields})
\begin{equation}\label{backgroundcond1}
\partial X^{(\mu)}_A / \partial s^{\sigma} = \delta^{(\mu)}_{\sigma} +\partial_{\sigma} \zeta^{(\mu)}_A(s) + \mathcal{O}(h^2),
\end{equation}
with   
$\partial_{\sigma} \zeta^{(\mu)}_A (s)\sim \mathcal{O}(h)$ being of the same order as the metric perturbation, and  $\zeta^{(\mu)}_A(s)$ need not be small in amplitude. 
Up to a constant irrelevant term, this gives
\begin{equation}\label{QCFexpand}
  X^{(\mu)}_A(s)= s^\mu  + \zeta^{(\mu)}_A(s).
\end{equation}
The first term is the background coordinate, while $\zeta^{(\mu)}_A(s)$ is the deviation of the \QCF{} from the value of the background coordinate chart $s$. The corresponding canonical momenta are therefore
\begin{equation}
\Pi^{(\mu)} = \Upsilon \partial_0 X^{(\mu)} =\Upsilon (\delta^{(\mu)}_{0} +\partial_{0} \zeta^{(\mu)} (s) + \mathcal{O}(h^2)).
\end{equation}
This indicates that
when we move to the quantum case, 
the momenta fluctuation of X-QRFs cannot be unbounded. Using operators smeared at a fixed spatial resolution with test function $f_s$, we require every  admissible state to satisfy the following bound on its momentum fluctuations: 

\begin{equation}\label{PIuncertaintyM}
\Delta \hat{\Pi}^{(\mu)}_{f_s}   \leq  \Upsilon  \mathcal{O}(h).
\end{equation}
This requirement is independent of the usual argument based on gravitational backreaction, but is consistent with the need to control matter momentum fluctuations in the weak-field regime. The uncertainty relation then gives, for every such state,
\begin{equation}\label{XuncertaintyM}
\Delta \hat{X}^{(\mu)}_{f_s}   \ge \frac{\hbar}{2  \Upsilon  \mathcal{O}(h) }    \int d^3\vec{u} f_s(\vec{u})^2,
\end{equation}
i.e. there is a lower bound on the uncertainty of the corresponding X-QRF. This means that the admissible X-QRF states cannot be arbitrarily sharply peaked on a definite field configuration.

Squeezed coherent states provide natural candidates for admissible  X-QRFs states.  Let us denote them as $\{ |\gamma^{\alpha}_{\mathrm{p}}  \rangle \}$, in which $\mathrm{p}:=(\mathrm{x,\pi})$ labels the expectation value of the field and momenta, and $\alpha$ labels the width of the squeezed coherent states. Both $\mathrm{p}$ and $\alpha$ can take on different values.
We require admissible states to satisfy  Eq.\,(\ref{PIuncertaintyM}, \ref{XuncertaintyM}) and the following conditions: 
\begin{itemize}[leftmargin=1 em]
\item  The expectation values of the smeared field operators on the admissible states are smooth and locally invertible functions, analogous to classical coordinates: 
\begin{equation}\label{meanfield}
\langle \gamma^{\alpha}| \hat{X}^{(\mu)}_{f_s}  |\gamma^{\alpha}\rangle = \int d^3\vec{u} f_s(\vec{u})\  x^{(\mu)} (s^0,\vec{u}) := \mathrm{x}_f^{(\mu)}(s).
\end{equation}

\item  The following bound is valid on admissible states:
\begin{equation}
||\partial_{\sigma} \hat{\zeta}^{(\mu)}_A |\gamma^{\alpha}\rangle ||\sim\mathcal{O} ( h),
\end{equation}
in which $||...||$ is the norm of the state. 

\item  The following bounds on the canonical momenta holds on admissible states:
\begin{equation}
|| (\hat{\Pi}^{(0)} - \Upsilon \hat{\mathbb{1}})|\gamma^{\alpha}\rangle || = \Upsilon \mathcal{O}(h) , || \hat{\Pi}^{(i)} |\gamma^{\alpha}\rangle ||\sim  \Upsilon \mathcal{O}(h). 
\end{equation}
\end{itemize}
The above conditions lead to an expansion of the \QCF{} stress-energy tensor on the admissible states: the leading term in $\hat{T}^{0i}_{A}$ is 
linear in the canonical momenta $\hat{\Pi}^{(i)}_{A}$:
\begin{equation}
\hat{T}^{0i}_{A} = \sum_{\mu}\hat{T}^{0i}_{A (\mu)}  = -\hat{\Pi}_A^{(i)} -\Upsilon \partial^i\hat{\zeta}_A^{(0)} + \mathcal{O}(\Upsilon h^2).
\end{equation}
Similarly, the energy density can be expanded as 
\begin{equation}
\hat{T}_A^{00} = \hat{\Pi}_A^{(0)} +f(\hat{X}_A^{(\mu)}) + \mathcal{O}(\Upsilon h^2).
\end{equation}
The expansions above, and similar ones throughout this work, should be understood as matrix elements between admissible states.  i.e. $\langle \gamma^{\alpha}_\mathrm{p}|....|\gamma^{\beta}_\mathrm{p'}  \rangle_A , \forall \alpha, \beta, \mathrm{p},\mathrm{p'}$ (see Appendix \ref{App:AdmiStates}); we write them skipping the admissible states to avoid overloading the notation.
As we see later, these properties are crucial for reducing to a \QCF{} perspective. With these conditions, the time evolution of the temporal component of the \QCF{}s takes the simple form $\hat{X}_A^{(0)}(s^{0} + t, \vec{s}) = \hat{X}_A^{(0)}(s^{0}, \vec{s}) + \hat{\mathbb{1}}_A\, t$.

In simple settings,  such as perturbations around a Minkowski background, the construction of $\hat{X}^{(\mu)}$ may be globally defined. In a generic spacetime, 
similarly to ordinary coordinate charts,  each set of \QCF{}s is expected to be valid only in a sufficiently small patch, while different patches in general require different \QCF{}s to be related by appropriate transition maps.

\section{Diffeomorphism-invariant relational observables} 

To eliminate the dependence on the unphysical manifold points, diffeomorphism-invariant observables can be constructed by describing a system relative to  dynamical, physical  fields~\cite{Rovelli1991,Rovelli:2001bz, Dittrich:2005kc, Dittrich2007,Tambornino:2011vg, Brunetti:2013maa, Giddings:2025xym, Baldazzi:2021fye,Goeller:2022rsx}. 
In the classical setting, consider  the observable $O$ evaluated where a  scalar field $\phi$ has value $y$. The composed quantity $O(\phi^{-1}(y))$ is then invariant under diffeomorphisms: although both $O$ and $\phi$ transform, their combination does not. Such quantities are also known as complete observables~\cite{Rovelli:2001bz, Dittrich:2005kc}. 
The dependence on the background manifold  is replaced by the dependence on the physical value  of a reference field, so that the observable is localized purely in relational terms~\cite{Goeller:2022rsx}.

When those reference fields are quantum, the joint relational observables between  a tensorial operator $\hat{F}_{\mu..}^{\nu..}(\vec{s})$  and   $\hat{X}_A$ can be constructed analogously the the classical case:
\begin{equation}\label{QDiracO}
\hat{F}_{\mu..}^{\nu..}(s) \in \mathcal{L} (\mathcal{H}_F ) \rightarrow 
\hat{F}_{\alpha..}^{\beta..} (\chi_A(\hat{\mathrm{Y}}_s)) \in \mathcal{L} (\mathcal{H}_F \otimes \mathcal{H}_A),
\end{equation}
in which $\hat{\mathrm{Y}}^{(\mu)}_s$ is  a set of non-dynamical operators defined as $\hat{X}^{(\mu)}_A(s)$ evaluated at an arbitrary specific coordinate point $s^\mu$.  The inverse operator $\mathbb{\chi}_A$, which is defined by the condition $\hat{X}^{(\mu)}_A ( \chi_A(\hat{\mathrm{Y}}_s) ) = \hat{\mathrm{Y}}_s$,  maps the \QCF{}  $\hat{X}_A$ back to the background spacetime manifold, see Appendix \ref{app:relob} for details.

\section{Reduction to the quantum reference field perspective}

\label{sec:T} 
The gauge-invariant physical state $\ket{\Psi}_{ph}$ encompasses all potential internal perspectives, but  not  described in any specific  internal  frame. 
In the QRF literature, this situation is captured by the  perspective neutral (PN) approach~\cite{Vanrietvelde:2018pgb, perspective2,delaHamette:2021oex}. This formulation of QRFs has two main elements: 1) an overarching description that does not correspond to the view from any physical frame (i.e.\,the PN structure), and 2) a reduction procedure to the perspective of a physical frame. The PN framework is obtained by enforcing some dynamical constraint(s) corresponding to the invariance of the physical systems under a group of transformations. In the $N$-particle case, for instance, one requires global translational and rotational invariance~\cite{Vanrietvelde:2018pgb, perspective2}. The reduction procedure consists in a unitary operator, called the trivialisation map and denoted as $\hat{\mathcal{T}}_A$, with $A$ being the QRF to which one wants to transform, and a subsequent projection that removes the redundant degree(s) of freedom resulting from the constraint(s) and that fixes the gauge. The PN framework has been developed for systems with symmetry or gauge groups when they are unimodular Lie groups \cite{delaHamette:2021oex}. Extending the framework to spacetime diffeomorphisms poses both technical and conceptual challenges. Beyond the fact that the diffeomorphism group is infinite-dimensional and not locally compact, meaning that the work of Ref.\,\cite{delaHamette:2021oex} cannot be directly applied, there is a conceptual reason related to the fact that the \QCF{}s are physical quantum fields. Specifically, due to universal coupling, any physical reference field has an associated stress-energy tensor, and therefore in principle back-reacts on the quantum state of the gravitational field. Thus, adding and changing the reference fields is not merely a passive relabelling of an otherwise fixed relational degrees of freedom; it also changes the physical system whose relational description is being constructed. Our construction is to address these challenges in the linearized regime.

Linearized quantum gravity is a first-class constrained system governed by Eq.\,\eqref{eq:FullHam} and Eq.\,\eqref{eq:constraintsLQG}. The physical Hilbert space of the theory satisfying the dynamical constraints provides the PN structure. To reduce to the description of a physical coordinate field, say $A$, we define the following constraint trivialisation map $\hat{\mathcal{T}}_{A}: \mathcal{H}_{PN} \rightarrow \mathcal{H}_{ABSG|A},$
\begin{equation}\label{TriMap}
\begin{split}
\!\! \hat{\mathcal{T}}_{A} (s^0) = &e^{-\frac{i}{\hbar}  \int d^3\vec{s}   \ \hat{\zeta}^{A}_{(\mu)}(s^0,\vec{s}) \hat{K}_{BSG}^\mu (s^0,\vec{s})} 
\end{split}
\end{equation} 
where
\begin{equation}
    \hat{K}_{BSG}^\mu (s^0,\vec{s})= \left(\hat{T}^{0\mu}_B+\hat{T}^{0\mu}_S +2\partial_{\nu} \hat{\pi}_G^{\nu \mu} +   \dfrac{\delta_0^\mu}{\kappa}\Delta \hat{h}^T\right)(s^0,\vec{s}).
\end{equation}
The operator $\hat{\zeta}^{(\mu)}_A$, defined in Eq.\eqref{backgroundcond1} and Eq.\eqref{QCFexpand}, denotes the difference between  the physical \QCF{} $\hat{X}^{(\mu)}_A$ and the background coordinates $s^{\mu}$, which was chosen arbitrarily. It can be obtained  from the \QCF{} operator $\hat{X}^{(\mu)}_A$ through a unitary field-translation operator,   $\hat{\zeta}^{(\mu)}_A (s)=\hat{U}_{X\rightarrow \zeta}(s^0)\hat{X}^{(\mu)}_A(s) \hat{U}_{X\rightarrow \zeta}^\dagger (s^0)$, where $\hat{U}_{X\rightarrow \zeta}(s^0) = e^{-\frac{i}{\hbar} \int d^3s' s'_\nu \hat{\Pi}^\nu_A(s')}$ and $s' = (s^0, \vec{s}')$.
With the trivilisation map acting on the constraints,  we obtain (see Appendix \ref{app:Tri} for details):
\begin{equation} \label{eq:trivialisationC}
\begin{split}
\hat{\mathcal{T}}_{A}  \hat{\mathcal{C}}^i (s) 	\hat{\mathcal{T}}^{\dagger}_{A} &=   \hat{T}^{0i}_{A|A}(s) + \mathcal{O}(h^2) \\
\hat{\mathcal{T}}_{A}  \hat{\mathcal{C}} (s) 	\hat{\mathcal{T}}^{\dagger}_{A} &=   \hat{T} ^{00}_{A|A} (s) + \mathcal{O}(h^2),
\end{split}
\end{equation}
where we used that, to leading order, the stress–energy tensors canonically commute with the coordinate fields, which is the consequence of the condition in Eq.\,\eqref{backgroundcond1} of \QCF{}s. 
Hence, the reduction to the \QCF{} $A$ yields a system in which, to our order of approximation, the momentum of $A$ is constrained to zero. The gauge fixing can be performed by projecting on the eigenvectors of the \QCF{} operators, $\ket{x}_A$, which is the canonically conjugated basis to the one diagonalizing the constraints.

The diffeomorphism invariant operators in Eq.\,\eqref{QDiracO} live on the PN (physical) Hilbert space. The operator $\hat{\mathcal{T}}_{A}$ maps such gauge-invariant operators to their  descriptions in the \QCF{} $A$. We now focus on observables of the type $\hat{O} (\chi_A(\hat{Y}))$.  For a  scalar field $\hat{\psi}_S$, the constraint trivialisation map yields, up to our order of approximation,
\begin{equation}
\hat{\mathcal{T}}_{A}  \hat{\psi}_S (\chi_A(\hat{Y}))	\hat{\mathcal{T}}^{\dagger}_{A} = \hat{\psi}_{S|A} (\hat{X}_{A|A}),
\end{equation}
i.e. the  scalar field expressed in $A$-coordinates (for a complete derivation, see Appendix~\ref{app:TMetric}).  
Such an operator can be expressed as an operator-valued function
\begin{equation}
\hat{\psi}_{S|A}(\hat{X}_{A|A}) := \int d\mu(x) \hat{\psi}_{S|A}(x) \otimes |x\rangle_{A|A} \langle x|, 
\end{equation}
acting on both perspectival Hilbert spaces of the scalar field $\mathcal{H}_{S|A}$ and the coordinate field  $\mathcal{H}_{A|A}$. Since a general physical state in the \QCF{} of $A$ has the form $|T_A^{0\mu}=0 \rangle_A \ket{\Psi}_{BSG|A}$ (see Appendix~\ref{app:States}), it is now clear that the gauge-fixing on the eigenvector $\ket{x}_A$ yields  
\begin{equation}
    {}_A\bra{x_A}\hat{\psi}_{S|A} (\hat{X}_{A})   | \psi \rangle_{S|A} |T_A^{0\mu}=0 \rangle_A =  \psi(x_A)   | \psi\rangle_{S|A},
\end{equation}
where $| \psi\rangle_{S|A}$ is the field basis of $\hat{\psi}_{S|A}$. Hence, we find that the reduction to the \QCF{} $A$ yields a representation of the relational observables in terms of the physical coordinates, intended as the eigenvalues of the \QCF{}. 
The case in which the field $\hat{\psi}$ is also a \QCF{} $\hat{X}_B^{(\mu)}$ is just a special case of a scalar field, namely  $\hat{\mathcal{T}}_A  \hat{X}^{(\mu)}_B(\chi_A(\hat{Y}_s)) \hat{\mathcal{T}}^{\dagger}_A  = \hat{X}^{(\mu)}_{B|A}(\hat{X}_{A})$.

The action of the constraint trivilisation map on the diffeomorphism invariant relational metric yields
\begin{equation}\label{PersMetric}
\hat{\mathcal{T}}_{A}  \hat{J}^{\mu}_{\alpha}  \hat{J}^{\nu}_{\beta}  g_{\mu \nu} (\chi_A(\hat{\mathrm{Y}}_s)) \hat{\mathcal{T}}_{A}^{\dagger} = \eta_{\alpha\beta}   + h_{\alpha\beta}(\chi_A(\hat{\mathrm{Y}}_s))  := g_{\alpha\beta}^{A},
\end{equation}
which corresponds to using  a particular \QCF{} $A$ to express the metric tensor. The $\alpha,\beta$ indices indicate that the description is in a specific frame, in this case of $A$.

\section{Perspectival quantum reference field transformation}

We now construct the transformations between \QCF{}s.
In the usual PN approach to QRFs, the transformation between the perspectives of $B$ and $A$ is  obtained  by  composing  the constraint-trivialisation maps that reduce to the two frames, i.e. $\hat{S}_{B\rightarrow A} = \hat{\mathcal{T}}_A \cdot \hat{\mathcal{T}}^{\dagger}_B$ \cite{Vanrietvelde:2018pgb, hoehn2021quantum}. 
This transformation can be generalized in a simple way: Before reducing to a particular \QCF{} perspective, one may act on the perspective-neutral Hilbert space with any unitary operator $\hat{V}$  that commutes with the constraints, e.g.\,a function of Dirac observables. It changes the properties of the physical state while not mapping it out of the constraint surface. This gives
\begin{equation}
\label{Sv}
\hat{S}^{V}_{A\rightarrow B} = \hat{\mathcal{T}}_A \hat{V} \hat{\mathcal{T}}_B^\dagger.
\end{equation}
For a non-trivial $\hat{V}$, this transformation is no longer a pure change of perspective. 
For example, in the quantum particle case~\cite{Vanrietvelde:2018pgb}, one may act with a unitary $\hat{V}$ to shift one particle relative to the others; 
this yields a system of particles that is still invariant under global translations, but  the relative positions between the particles are different. If $\hat{V}$ is chosen to be the  time evolution operator, then $\hat{S}^{V}_{B\rightarrow A}$ describes the evolution  as viewed through a change of perspective from $B$ to $A$. 

In particular, when $\hat{V} = \hat{\mathbb{1}}$, the resulting  \QCF{} transformation is purely a  change of perspective: it changes the physical coordinate field relative to which the system is described, while leaving the underlying physical configuration unchanged. It is in analogous to a coordinate transformation, but with the coordinates being physical fields.  
Starting from the perspective of B, $\hat{S}_{B\rightarrow A} (s^0) : =  \hat{\mathcal{T}}_{A}  \cdot  \hat{\mathcal{T}}^{\dagger}_{B}$ can be cast in the equivalent form as (see Appendix \ref{app:Trans})
\begin{equation}\label{perstran}
\begin{split}
&\!\!\!\hat{S}_{B\rightarrow A} (s^0)   = \hat{\mathcal{I}}_{BA}\cdot e^{-\frac{i}{\hbar} \int d^3s  \hat{\zeta}^{A|B}_{(\mu)}(s) \hat{T}^{0\mu}_{B|B}(s)} e^{\frac{i}{\hbar} \int d^3s  \hat{\zeta}^{B|B}_{(\mu)}(s) \hat{T}^{0\mu}_{A|B}(s)  } \\ 
&\cdot e^{-\frac{i}{\hbar}  \int d^3s   \hat{\zeta}^{A|B}_{(\mu)}(s) \left(\hat{T}^{0\mu}_{S|B}(s) +2\partial_{\nu} \hat{\pi}_{G|B}^{\nu \mu} (s) + \delta_0^\mu  \frac{1}{\kappa}\Delta \hat{h}^T_{G|B}(s)\right)} .
\end{split}
\end{equation}
in which $\hat{\mathcal{I}}_{BA}$ is a Hilbert space isomorphism that maps $\mathcal{H}_{..|B}$ to $\mathcal{H}_{..|A}$.
Therefore, a \QCF{} perpectival transformation is a \emph{quantum-controlled unitary transformation}. Its action on all systems not behaving as \QCF{}s is controlled by the relative displacement between the two \QCF{} $\hat{\zeta}^{A|B}_{(\mu)}(s)$, encoding local relational information specifying the transformation.   The structure of the transformation is analogous to \cite{Giacomini:2017zju}, but as it is a local field transformation, here we cannot remove the degree of freedom of the \QCF{}.

\begin{figure*}
\centerline{\includegraphics[width=0.9\textwidth]{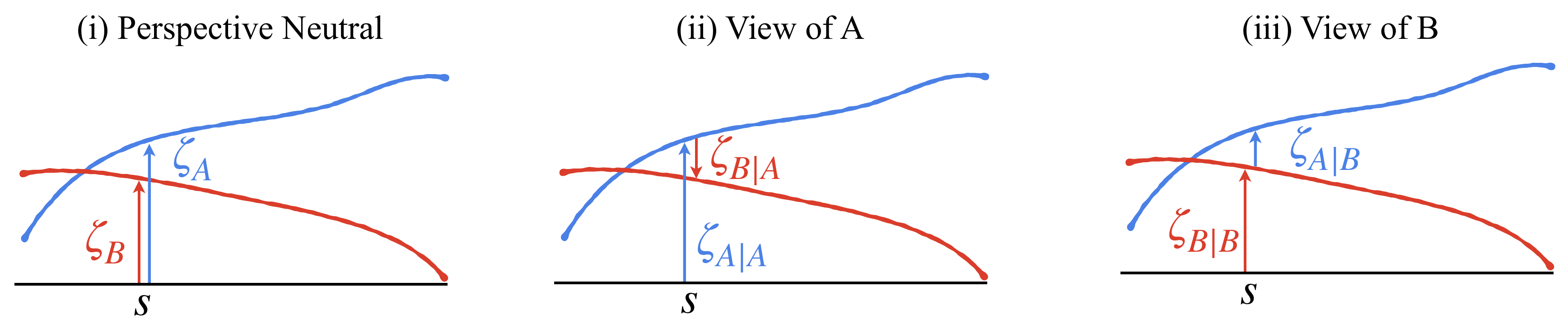}}     \caption{\label{fig:zetainpersp} 
The eigenvalues of the quantum fields $\hat{\zeta}_A(s)$ and $\hat{\zeta}_B(s)$ describing the difference between the \QCF{}s $\hat{X}_A(s)$ and $\hat{X}_B (s)$ and the arbitrarily chosen background coordinates $s^{\mu}$ in (i) the perspective neutral description, (ii) the \QCF{} of $A$, and (iii) the \QCF{} of $B$.}
\end{figure*}

From Eq.\,\eqref{perstran} it is easy to check that the action on the difference between the \QCF{}s and the background coordinates is (See Fig.\,\ref{fig:zetainpersp})
\begin{equation}
    \begin{split}
        &\hat{S}_{B\rightarrow A} \hat{\zeta}_{A|B} \left(\hat{S}_{B\rightarrow A}\right)^\dagger = - \hat{\zeta}_{B|A},\\
        &\hat{S}_{B\rightarrow A} \hat{\zeta}_{B|B} \left(\hat{S}_{B\rightarrow A}\right)^\dagger =  \hat{\zeta}_{A|A}+ \hat{\zeta}_{B|A}.
    \end{split}
\end{equation}
Finally, the transformation of the linearized gravitational field is
\begin{equation} \label{eq:transfhnonrel}
\hat{S}_{B\rightarrow A} \hat{h}_{\mu\nu}^{G|B}\left(\hat{S}_{B\rightarrow A}\right)^\dagger = \hat{h}_{\mu\nu}^{G|A} - \partial_{\nu} \hat{\zeta}^{B|A}_{(\mu)} - \partial_{\mu} \hat{\zeta}^{B|A}_{(\nu)}.
\end{equation}
 The transformation $\hat{S}_{B\rightarrow A}$   promotes the vector fields in the  linearized diffeomorphisms  to a quantum field with a physical origin: the relative displacement between two \QCF{}s. 
 Importantly, this is not a gauge transformation, but a physical perspectival transformation between \QCF{}s that extends the classical notion of coordinate changes. Its action on $\hat{h}_{\mu \nu}$ structurally mirrors the standard linearized diffeomorphism, but cannot be obtained within the standard treatment, where the transformation parameter is classical gauge parameter. 
Such quantum extension is necessary once the metric perturbation itself is a quantum operator, ensuring that $\hat{h}_{\mu\nu}$  transforms consistently according to an  operator equation.

 Eq.\,\eqref{eq:transfhnonrel} immediately implies that the Hilbert space of the gravitational field and the Hilbert space of the matter fields are not fixed \emph{a priori}, but depend on the perspective that we take to define the physics. This is a general feature that arises in most works on QRFs~\cite{Giacomini:2017zju,Vanrietvelde:2018pgb, de2020quantum}, and we comment on it in the context of linearized gravity in Section~\ref{sec:Discussion}.

The \QCF{} transformation of the relational observables, for instance for a scalar field, yields 
\begin{equation}
\hat{S}_{B\rightarrow A} \hat{\psi}( \hat{X}_{B|B})(\hat{S}_{B\rightarrow A})^{\dagger}  = \hat{\psi}(\hat{X}_{A|A}),
\end{equation}
For the perspectival metric operator $\hat{g}_{\alpha\beta}^B $ defined in Eq.\eqref{PersMetric}, we  obtain the description in  A's perspective (see Appendix \ref{app:Trans})
\begin{equation}\label{hpitrans}
\begin{split}
& \hat{S}_{B\rightarrow A} \cdot   \hat{g}_{\alpha\beta}^B \cdot  (\hat{S}_{B\rightarrow A})^{\dagger} \\
=\ & \hat{g}_{\alpha \beta}^A -\partial_{\alpha}\hat{\zeta}_{(\beta)}^{B|A} (\hat{X}_{A|A})-\partial_{\beta}\hat{\zeta}_{(\alpha)}^{B|A} (\hat{X}_{A|A}),
 \end{split}
\end{equation}
which realizes a local quantum coordinate transformation as a \QCF{} transformation.

\section{Diffeomorphisms in a quantum reference field perspective}
\label{CompareDiff}

In general relativity, passive diffeomorphisms re-express the same tensor field in different coordinates, whereas active diffeomorphisms  map  the manifold to itself, thereby generating new tensor fields.
In our work, the \QCF{} transformations are physically and conceptually distinct from passive diffeomorphisms, although they may look similar to them.
This is because,  the \QCF{}s are physical fields within the full system, active diffeomorphisms act on them just as they do on all other fields. Let us consider a one-parameter group of diffeomorphisms generated by a vector field  $l^{\mu}$. In a specific \QCF{} perspective, it is straightforward to derive  the reduced form of the standard (active) diffeomorphism.
Upon applying the map $\hat{\mathcal{T}}_{A}$, we obtain the active diffeomorphism transformation in the perspective of A:
\begin{equation} \label{reduceactive}
\hat{S}^a_A  := \hat{\mathcal{T}}_{A}  e^{-\frac{i}{\hbar}  \lambda \hat{G}_{l}(t)}  \hat{\mathcal{T}}^{\dagger}_{A}=  e^{-\frac{i}{\hbar}  \int d^3s \lambda  l^{\mu}(s) \hat{T}_{0\mu}^{A|A}(s) + \frac{i}{\hbar} \mathcal{O}(h^2)}  
\end{equation}
in which $\hat{G}_l$ is generators of diffeomorphisms along vector field $l^\mu$ in Eq.\eqref{Main:fullgen}. The perspectival observables in the previous sections are invariant under the above transformation, due to the fact that they are constructed from the relational operator at PN level $\psi(\chi_A(\hat{\mathrm{Y}}_s))$, $ e^{-\frac{i}{\hbar}  \lambda \hat{G}_{l}(t)} \psi(\chi_A(\hat{\mathrm{Y}}_s)) e^{\frac{i}{\hbar}  \lambda \hat{G}_{\hat{l}}(t)} = \psi(\chi_A(\hat{\mathrm{Y}}_s))$, For perspectival observables,  we have
\begin{equation} \begin{split}
\hat{S}^a_A \psi_{S|A}(\hat{X}_{A|A}) (\hat{S}^a_A)^{\dagger} & = \hat{\mathcal{T}}_{A} e^{-\frac{i}{\hbar}  \lambda \hat{G}_{l}(t)} \psi(\chi_A(\hat{\mathrm{Y}}_s)) e^{\frac{i}{\hbar}  \lambda \hat{G}_{l}(t)} \hat{\mathcal{T}}^{\dagger}_{A}\\
& =  \psi_{S|A}(\hat{X}_{A|A}).
\end{split}
\end{equation}
Hence, the active diffeomorphisms act as the identity operator in the reduced Hilbert space relative to a \QCF{}, as expected. The mathematically equivalent passive transformation is a  change of coordinate chart $s^{\mu}$. This is different from the \QCF{} transformations $\hat{S}_{B\rightarrow A}$, which implements a change of perspective between descriptions defined with respect to two quantum fields.

\section{Operational considerations on relational observables}

We now show that the reduced relational observables constructed above can, at least in principle, be accessed operationally from the perspective of an X-QRF.
As we detail in Appendix~\ref{app:vNeum}, consider a system $S$ and a probe $P$ which are both scalar fields, and a quantum clock $C$, all treated as physical degrees of freedom entering the gravitational constraints. At the perspective-neutral level, the measurement interaction is built from diffeomorphism-invariant relational observables.
After reducing to the perspective of a \QCF{} $A$, the interaction Hamiltonian becomes
\begin{widetext}
\begin{equation} \label{eq:HamIntMeasPsi}
\begin{split}
\hat{H}_{int|A}= \int d^3\vec{s}\, \delta\left(\hat{\tau}_{C}(\hat{X}^{(0)}_{A|A}(s))-t^* \hat{\mathbb{1}}\right) \hat{\psi}_{S|A} (\hat{X}_{A|A}(s)) \hat{\phi}_{P|A}(\hat{X}_{A|A}(s)).
\end{split}
\end{equation}
\end{widetext}
The interaction is activated when the quantum clock $\hat{\tau}_{C|A}$ has eigenvalue $\tau_{C|A} = t^*$. Consequently, when $\hat{H}_{int|A}$ acts on a generic quantum clock state, the interaction is triggered at different values of $x^0_A$ depending on the clock eigenstate, similarly to Refs.\,\cite{Zych:2017tau, castro2020quantum, giacomini2021spacetime}. 
After the interaction, conditioning on the clock reading and measuring the probe gives a conditional probability to measure an outcome $n$ given that the clock shows time $\tau$, namely
\begin{equation}
\begin{split}
p_{n|\tau} &= \frac{\mathrm{Tr}_{SCP}\left[\ket{\Phi_t}\bra{\Phi_t}\left(\hat{\Pi}_n^{P|A} \otimes \hat{\mathbb{1}}_{S|A}\otimes \hat{\Pi}_{\tau}^{C|A}  \right) \right]}{\mathrm{Tr}_{SCP}\left[\ket{\Phi_t}\bra{\Phi_t}\left(\hat{\mathbb{1}}^{P|A} \otimes \hat{\mathbb{1}}_{S|A}\otimes \hat{\Pi}_{\tau}^{C|A} \right) \right]},\\
\end{split}
\end{equation}
where $\ket{\Phi_t}$ is the state of the system, the probe, and the clock after the interaction
\begin{equation}
    \ket{\Phi_t}_{|A} = \overleftarrow{\mathrm{T}}\left( e^{-\frac{i}{\hbar}\int_0^t dx_A^0 \hat{H}_{int|A} (x_A^0)}\right) \ket{\Phi_0}_{|A},
\end{equation}
and $\ket{\Phi_0}$ is the initial state, prepared as a product state of the three systems. In addition, $\hat{\Pi}_n^{P|A}$ and $\hat{\Pi}_{\tau}^{C|A}$ are projectors over an eigenvalue of the momentum of the probe and of the time reading of the clock respectively. The coupling shifts the pointer by an amount determined by the eigenvalue of the measured relational observable $\hat{\psi}_{S|A}(\hat{X}_{A|A})$ (see again Appendix~\ref{app:vNeum} for details).

Crucially, as the coupling and projector are constructed from diffeomorphism-invariant relational observables, this ensures that the measurement is consistent with the gauge constraints of the theory. 
The conditioning (gauge-fixing) on a specific choice of coordinates of the \QCF{} $A$ allows us to express all quantities in the eigenvalues of $\hat{X}^\mu_{A|A}(s)$, thus providing a physical coordinate system on which the measurement interaction takes place. Thanks to that, we remove any reference to the background coordinate system $s^\mu$. We demonstrate the idea through the example of a scalar field observable in Eq.\,\eqref{eq:HamIntMeasPsi}, but the construction extends naturally to relational gravitational observables. The operational implications, as well as more sophisticated and formal development of relativistic relational measurement theory, for instance by using the techniques in  \cite{FewsterVerch2020,Fewster:2024pur, Simmons:2025cmv}, will be addressed in the future work.   

\section{Discussion}
\label{sec:Discussion}

In this work, we constructed  quantum reference fields (\QCF{}s), defined as  physical scalar fields whose stress-energy tensors enter the gravitational constraints, and we showed that they provide a relational description of linearized quantum gravity with respect to internal matter subsystems. 
By generalizing the perspective-neutral construction of QRFs to linearized quantum gravity, we constructed a general class of transformations between \QCF{}s, which in general change the physical configuration, and identified relational observables and physical states in the perspective of each \QCF{}. Among all possible \QCF{} transformations,
those that preserve the physical configuration, i.e. those for which $\hat{V} = \hat{\mathbb{1}}$ in Eq. \eqref{Sv}, provide genuinely quantum analogues of coordinate transformations. Crucially, however, they are conceptually and physically distinct from diffeomorphisms, or superpositions of diffeomorphisms \cite{Kabel:2024lzr}. In Section \ref{CompareDiff}, we show that the perspectival observables transform nontrivially under \QCF{} transformations, while remaining invariant under diffeomorphisms.
Finally, we outlined a von Neumann measurement scheme for relational observables, arguing that perspectival observables are, in principle, operationally accessible.

The features discussed in this work have two distinct origins: first, the fact that the references are dynamical subsystems rather than external structures, a feature already relevant when the \QCF{}s are treated classically; and second, the genuinely quantum nature of \QCF{}s.

On the one hand, the structure of the transformations is due to the dynamical subsystem  nature of the fields, as opposed to fictitious coordinates or vector fields giving rise to the diffeomorphism transformations in general relativity. Promoting the parameter of the transformation to a physical, dynamical variable could by itself give rise to an algebra of the transformations that differs from the usual ones. This occurs, for instance, in the classical limit of the frames in  Galilean quantum reference frames, where the dynamical nature of the reference frames leads to a group of inertial transformations that is different to the centrally-extended Galilei group (see Ref.~\cite{ballesteros2020group}). On the other hand, some of the features we discussed here have no classical counterpart. These include the possibility of performing a quantum superposition of different transformations, the quantum dynamics of the \QCF{}s, and the preservation of the constraint structure by the post-measurement state when relational observables are measured within a von Neumann measurement scheme.

A fundamental feature  revealed by the construction is that the distinction into ``gravity'' and ``matter'' degrees of freedom is not absolute, but depends on the specific perspective chosen. From Eq.\,(\ref{hpitrans}), we can see the explicit mixing of the Hilbert spaces of the \QCF{}s, $\mathcal{H}_A$ and $\mathcal{H}_B$, with the Hilbert space of perturbative quantum gravity, $\mathcal{H}_G$. This places the matter and gravitational sectors on a more equal footing in the linearized regime, and suggests that such feature may extend to the non-perturbative regime.  The dependence of the subsystem structure on the quantum reference frame is also present in the single particle case (see, e.g.\,\cite{Giacomini:2017zju, Vanrietvelde:2018pgb, hoehn2021quantum}). In gravity, however, the implication is far more radical: the usual separation between gravitational degrees of freedom, which define spacetime structure, and matter degrees of freedom, which are taken to live on and curve that spacetime, ceases to be absolute, but becomes QRF-dependent. 

The possibility of changing between \QCF{}s through local transformations naturally raises the question of which gravitational configurations can be mapped one into another. 
Although certain superpositions of metric configurations can be transformed, through QRF changes of perspective, into a definite one \cite{de2021falling, Kabel:2022cje, delaHamette:2022cka}, the \QCF{}s transformations constructed here cannot map a genuinely quantum gravitational field to  a semiclassical one.  Since \QCF{} transformations are unitary, they preserve the canonical commutator of both the matter source and the gravitational field: they transform the perspective, but not its underlying quantum nature.  A natural line of research is to construct \QCF{}s from non-ideal reference frames; in that case, it may be possible that a quantum configuration of gravitational field effectively appears semiclassical for certain types of non-ideal frames.

The present work opens up many directions for future research. These include identifying new physical effects that arise directly from the quantum nature of the reference fields, investigating the notion of covariance in quantum gravity associated with  \QCF{}s transformations, utilising \QCF{}s in  the formulation of observers in quantum gravity, connecting the construction to edge modes as reference frames associated with boundary of spacetime \cite{{Donnelly:2016auv, Carrozza:2022xut, Kabel:2023jve, Janssen:2025uzf, fewster2025semilocalobservablesedgemodes}}, developing quantum error-correction perspectives on linearized gravity via QRFs \cite{Carrozza:2024smc,Rothlin:2026fqy,Lacambra:2026twp},  and ultimately extending the framework beyond the perturbative regime.

\raggedbottom

\acknowledgments  
We would like to acknowledge \v{C}. Brukner, G. Barnich, F. Dowker, M. Fahn, R. Falcone, K. Giesel, S. Garmier, P. Höhn, V. Kabel,  L. Loveridge, R. Loll, D. Oriti, P. Pulakkat, S. Popescu,  A. Roura, A. Smirne, R. Simmons,  R. Renner, E. Rothlin, A. Wall, R. Wald and G. Yang for inspiring and helpful discussions.  L.Q.C would like to thank the discussion and feedback from Brukner's group meeting, and the especially clarifying discussion with F. Dowker and D. Oriti, on the distinction between QRF, symmetry and gauge transformation during the Bratislava conference. 
L.Q.C is funded by Österreichischer Wissenschaftsfonds FWF ESPRIT fellowship with grant DOI: 10.55776/ESP390. L.Q.C thanks the Quantum Information Theory (QIT) group at ETH Zürich for  hospitality during her visit, where part of this work was carried out. She acknowledges support from the NCCR SwissMAP for the ETH visits. F.G. acknowledges support from the Swiss National Science Foundation via the Ambizione Grant PZ00P2-208885 and from the ETH Zurich Quantum Center.
This work was made possible through the support of the WOST, \href{https://withoutspacetime.org}{WithOut SpaceTime} project, supported by Grant ID\#~63683 from the John Templeton Foundation (JTF) and of the \href{https://www.templeton.org/grant/the-quantuminformation-structure-ofspacetime-qiss-second-phase}{QISS 2} ‘The Quantum Information Structure of Spacetime, Second Phase’ Project,  Grant ID\#~62312 from the John Templeton Foundation. The opinions expressed in this work are those of the author(s) and do not necessarily reflect the views of the John Templeton Foundation. 
This work was carried out with the support of the Italian Ministry of University and Research (MUR) under the Funding for Research Projects (FIS 2), pursuant to Ministerial Decree No. 23314 of 11/12/2024 (Project Q-GraSp, FIS-2023-02629).
F.G. is grateful for the hospitality of Perimeter Institute where part of this work was carried out. Research at Perimeter Institute is supported in part by the Government of Canada through the Department of Innovation, Science and Economic Development Canada and by the province of Ontario through the Ministry of Economic Development, Job Creation and Trade. This research was also supported in part by the Simons Foundation through the Simons Foundation Emmy Noether Fellows Program at Perimeter Institute. This work contributes to the European Union COST Action CA23130 \textit{Bridging high and low energies in search of quantum gravity} and has benefited from the activities of COST Action CA23115: \emph{Relativistic Quantum Information}, funded  by COST (European Cooperation in Science and Technology).

 \appendix

 \begin{widetext}

\section{Linearized Gravity and the Gauge Generators in the Enlarged Phase Space}
\label{app:Enlarged}
In the Arnowitt-Deser-Misner (ADM) formulation, one performs a 3+1 decomposition of the spacetime metric, i.e.~$ds^2 = - N^2 dt^2 + \gamma_{ij} (dx^i + N^i dt) (dx^j + N^j dt)$, with $i,j=1,2,3$. Here,  $\gamma_{ij}$ is the metric on the space-like foliation and $N$ and $N^i $  are the lapse function and shift vector, respectively. 
The  Hamiltonian of pure general relativity, without matter sources, reads \cite{PhysRev.116.1322,Thiemann2007}:
\begin{equation}
H_{ADM}= \int d^3s\left( \lambda(\vec{s}) P_0 (\vec{s}) +  \lambda^a (\vec{s}) P_a (\vec{s})  + N (\vec{s}) H (\vec{s}) +  N^a (\vec{s}) H_a (\vec{s}) \right) := P_0 [\lambda]   + \vec{P} [\vec{\lambda}] +  H [N] + \vec{H}[\vec{N}].
\end{equation}
The canonically conjugated momenta to the  lapse function $N$ and the shift vector $N_i$, respectively $P_0$ and $P_i$, are primary constraints, and hence vanish on the constraint surface. However, one can still work with them on an enlarged gravitational phase space and reduce the redundancy later. The generators of (infinitesimal) diffeomorphisms on such an enlarged phase space  are defined as \cite{Pons:1996av}
\begin{equation}\label{fullgen}
G_{\zeta}(t) =  P_{\mu} \dot{\zeta}^{\mu} + (\mathcal{H}_{\mu} + N^{\rho} K^{\nu}_{\mu \rho} P_{\nu}) \zeta^{\mu}
\end{equation}
which can be written more explicitly as
\begin{equation}\label{fullgen}
G_{\zeta}(t) =  \int d^3s \left( P_{\mu} (\vec{s})\dot{ \xi}^{\mu} (t, \vec{s}) + \mathcal{H}_{\mu} (\vec{s})  \xi^{\mu} (t, \vec{s}) \right)  +  \int d^3x  d^3y  \int d^3s  N^{\rho} (\vec{x}) K^{\nu}_{\mu \rho} (\vec{x},\vec{y};\vec{s}) P_{\nu}(\vec{s})  \xi^{\mu}(t, \vec{y}),
\end{equation}
where $ \xi^\mu(t, \vec{s})$ are arbitrary time-dependent functions that parametrize the diffeomorphism transformation. In addition, the structure functions $C^{\nu}_{\mu \rho}$ are defined through $\{\mathcal{H}_{\mu},\mathcal{H}_{\nu}\}= C^{\sigma}_{\mu \nu} \mathcal{H}_{\sigma}$. One can obtain the structure functions from the hypersurface deformation algebra~\cite{Thiemann2007}: 
\begin{equation}
\begin{split}
\{H [N], H [M]\} &= H_i \left[\gamma^{ij}(N\partial_j M - M\partial_j N )\right]  \\
\{\vec{H} [\vec{N}], H [M]\} &= H \left[\mathcal{L}_{\vec{N}}  M \right]  = H \left[N^i\partial_i M \right]  \\
\{\vec{H} [\vec{N}], \vec{H} [\vec{M}]\} &= \vec{H} \left[\mathcal{L}_{\vec{N}}  \vec{M} \right]  \\
\end{split}
\end{equation}
Therefore, the structure functions are differential operators
\begin{equation}
\begin{split}
C^{i}_{00} (x,y;z) &=   \gamma^{ij} \left(\delta (z-x)  \partial_j^{(z)} \delta (z-y)- \delta (z-y)  \partial_j^{(z)} \delta (z-x)\right), \\
C^{0}_{0i} (x,y;z) &=    \delta (z-x)  \partial_i^{(z)}\delta (z-y), \\
C^{k}_{ij} (x,y;z) &=   \delta^k_i \delta (z-y) \partial_j^{(z)} \delta (z-x)   -\delta^k_j \delta (z-x)  \partial_i^{(z)} \delta (z-y).
\end{split}
\end{equation}
Note that only $C^{i}_{00}$ depends on the metric of the spatial hypersurface.

In this work, we consider the first order of the perturbations around the flat metric, i.e.\,$g_{\mu\nu} = \eta_{\mu\nu} +  h_{\mu\nu}$, with $|h_{\mu\nu}| \ll 1.$ The perturbation in the 3+1 decomposition is expressed as	$\gamma_{ij} = \delta_{ij} +  h_{ij},\  N = 1 +n,\ N^i =0+n^i $. The canonical momenta and the linearized gravitational field satisfy the Poisson brackets: $\{ h_{ij} (\vec{x}), \pi^{kl} (\vec{x'})\} = \delta^k_{(i} \delta^l_{j)} \delta^3(\vec{x}-\vec{x}') $, with $ \pi^{kl} = \frac{1}{2\kappa} ( \dot{h}^{kl} - \dot{h} \delta^{kl} ) $.  We enlarge the phase space to include $p^{\mu}$, i.e.\,the canonical conjugate momenta to $n$ and $n_i$ respectively.  The total Hamiltonian of the linearized gravitational field together with other matter content S in the system is
\begin{equation} 
\hat{H}_{tot} =  H_S +  H_G   -\frac{1}{2}\int d^3s \hat{h}_{ij} (\vec{s})T^{ij} (\vec{s}) 
    + \int d^3s \,\left(n(\vec{s}) \mathcal{C} (\vec{s})+ n_i (\vec{s}) \mathcal{C}^i(\vec{s})  + \lambda_{\mu} (\vec{s}) \pi^{0\mu} (\vec{s})  \right),
\end{equation}
in which $ H_G$ is a quadratic Hamiltonian obtained from expanding $H_{ADM}$ to the quadratic order on the Minkowski background, with $\kappa = 16\pi G c^{-4}$:
\begin{equation}
H_{G}  = \kappa  \int d^3x  \  \left(\pi_{kl} \pi^{kl} - \pi^2/2 \right) + \frac{1}{4\kappa}   \int d^3x [ \partial_k  h_{ij} \partial^k  h^{ij}  - \partial_i h \partial^i h - 2 \partial^k h_{ik} (\partial_j h^{ij}-\partial^i h) ].
\end{equation}
The four constraints are
\begin{equation} 
    \begin{split}
        &\hat{\mathcal{C}}(\vec{s}) : =  \partial_i \partial^i \hat{h}^T(\vec{s}) + \kappa \hat{T}^{00}(\vec{s}), \\
        &\hat{\mathcal{C}}^i(\vec{s}) : = 2 \partial_j \hat{\pi}_G^{ij} (\vec{s})+ \hat{T}^{0 i}(\vec{s}) ,
    \end{split}
\end{equation}
where $h_T= \delta^{ij} h_T^{ij}$, $h_T^{ij}$ is the transverse part of the metric perturbation $ h_T^{ij} (\vec{k}):=P^i_k P^j_l h^{kl} (\vec{k}) $ and $P^i_j = \delta^i_j - \frac{k^i k_j}{|\vec k|^2}$ is the projector along the transverse direction to the momentum $\vec{k}$.

Evaluating the hypersurface deformation algebra on the Minkowski background, one obtains the $K^{i}_{00}$ component of the structure function
\begin{equation}\label{Ki00}
\tilde{K}^{i}_{00}(x,y;z):=K^{i}_{00}|_{\eta}(x,y;z) =   \delta^{ij} \left(\delta (z-x)  \partial_j^{(z)} \delta (z-y)- \delta (z-y)  \partial_j^{(z)} \delta (z-x)\right).
\end{equation}
The generators of diffeomorphisms on the enlarged phase space of linearized gravity can be written as
\begin{equation}
G_{\xi}(t) = \int d^3s  \left( p_\mu(\vec{s}) \dot{\xi}^{\mu}(\vec{s}) + \mathcal{C}_\mu(\vec{s}) \xi^{\mu} (\vec{z}) \right) + \int d^3x  d^3y \int d^3z  \tilde{K}^{\nu}_{\mu0} (x,y;z)  \xi^{\mu}(\vec{y}) p_{\nu} (\vec{z}).
\end{equation}
Note that the leading order of $N^i$ vanishes, so only the term of $\tilde{K}^{\nu}_{\mu0}$ contributes to the generator.
We can readily check that 
\begin{equation}
\begin{split}
\delta_{\xi} n (\vec{s}) =\{G_{\xi}(t), n (\vec{s})\} &= \dot{\xi}^0 (\vec{s}) + \int d^3x  d^3y \int d^3z  \tilde{K}^{\nu}_{\mu0} (x,y;z)  \xi^{\mu}(\vec{y}) \delta^0_{\nu} \delta (\vec{z}-\vec{s})  \\
& = \dot{\xi}^0 (\vec{s}) + \int d^3x  d^3y  \delta(\vec{s}-\vec{y}) \partial^{(s)}_i \delta (\vec{s}-\vec{x}) \xi^{i}(\vec{y}) \\
& = \dot{\xi}^0 (\vec{s}),
\end{split}
\end{equation}
in which the second term vanishes in the second line. 
We also have
\begin{equation}
\begin{split}
\delta_{\xi} n^i (\vec{s}) =\{G_{\xi}(t), n^i (\vec{s})\} &= \dot{\xi}^i (\vec{s}) + \int d^3x  d^3y \int d^3z  \tilde{K}^{i}_{00} (x,y;z)  \xi^{0}(\vec{y})  \delta (\vec{z}-\vec{s})  \\
&= \dot{\xi}^i (\vec{s})+ \partial^i \xi^0(\vec{s}) 
\end{split}
\end{equation}
where we have used the nontrivial component of the structure function in Eq.\,(\ref{Ki00}) evaluated on the Minkowski background. 
Therefore, we have
\begin{equation}
\begin{split}
\delta_{\xi} n (\vec{s}) &=\{G_{\xi}(t), n (\vec{s})\} = \dot{\xi}^0 (\vec{s}), \\
\delta_{\xi} n^i (\vec{s}) &=\{G_{\xi}(t), n^i (\vec{s})\} = \dot{\xi}^i (\vec{s})+ \partial^i \xi^0(\vec{s}),\\
\delta_{\xi} h_{ij} (\vec{s}) &=\{G_{\xi}(t), h_{ij} (\vec{s})\} = 2 \partial_{(i} \xi_{j)}(\vec{s}).
\end{split}
\end{equation}

Crucially, the background structure functions of the full ADM Dirac algebra, evaluated on the Minkowski background, must be incorporated into the linearized gauge generator on the enlarged phase space in order to reproduce the correct classical diffeomorphism transformations of $h_{0\mu}$.

\section{Classical Conditions on the Quantum Reference Fields in the Perturbative Regime}
\label{app:CoordFields}

The perturbation of the gravitational field is defined with respect to the Minkowski background $\eta_{\rho\sigma}$, i.e.  $ g_{\rho\sigma} =  \eta_{\rho\sigma} + h_{\rho\sigma}$. In this  setting, the coordinates are not arbitrary; they need to satisfy certain conditions to preserve the Minkowski background plus perturbation. In this Appendix, we derive these conditions in the classical case, and then analyze the stress-energy tensor, when the  coordinates are physical fields. The corresponding quantum expressions can be obtained by directly quantizing the classical ones, with the caveat that the classical conditions translate as specific operator conditions on the set of admissible states (see Section~\ref{sec:admissibleXQRF} and Appendix~\ref{App:AdmiStates}).
Concretely, the transformation from the background coordinates $s^\mu$ to the coordinates $X^{(\mu)}$  yields
\begin{equation}
\frac{\partial s^\rho}{\partial X^{(\mu)}}\frac{\partial s^\sigma}{\partial X^{(\nu)}} g_{\rho\sigma}  = \frac{\partial s^\rho}{\partial X^{(\mu)}}\frac{\partial s^\sigma}{\partial X^{(\nu)}}  \left( \eta_{\rho\sigma} + h_{\rho\sigma}\right)\overset{!}{=} \eta_{\mu\nu}+ \tilde{h}_{\mu\nu}, 
\end{equation}
where in the last step we have imposed that the Minkowski background is preserved by the transformation. The previous condition is satisfied if the Jacobian respects the following expansion
\begin{equation}\label{backgroundcond}
\partial X^{(\mu)} / \partial s^{\sigma} = \delta^{(\mu)}_{\sigma} +\partial_{\sigma} \zeta^{(\mu)} (s) + \mathcal{O}(h^2),
\end{equation}
with $\partial_{\sigma} \zeta^{(\mu)} (s) = \mathcal{O}(h)$,  of the same order as the metric perturbation.  Therefore, to the order of consideration, the transformation to $X^{(\mu)}$ is a linearized diffeomorphism, namely 
\begin{equation}\label{XS}
X^{(\mu)} (s) = s^{(\mu)} + \zeta^{(\mu)} (s),
\end{equation}
with $\mu = 0,1,2,3$. 
 Here, $\zeta^{(\mu)} (s)$ is an arbitrary smooth  function. 
Note, however, we do not need to require $\zeta^{(\mu)} (s)$ itself to be small. Since $X^{(\mu)}$ serves as coordinate field in the construction, we take it to have dimension $[L]$ to be consistent with coordinates. In the scalar field Lagrangian, the kinematical terms  have a  parameter $\Upsilon$, which has dimension $[L^{-4}]$: 
\begin{equation}
\mathcal{L}_A= - \frac{\Upsilon}{2} \partial_\alpha X \partial^\alpha X  - V(X)
\end{equation}
with $X:= X^{(\mu)}_A$. 
 The stress-energy tensors are
\begin{equation}
T_{\alpha \beta}^{A (\mu)}=    \Upsilon \partial_\alpha X \partial_\beta X - \eta_{\alpha \beta} \left(\frac{\Upsilon}{2}  \partial_\rho X \partial^\rho X  + V(X)\right). 
\end{equation}
 
 Eq.\,(\ref{backgroundcond}) constrains  the  four physical scalar fields $ X^{(\mu)}$. From the scalar field Hamiltonian, their canonical momenta are $\Pi^{(\mu)} =\Upsilon \partial_0 X^{(\mu)}$.  In canonical variables, the stress-energy tensors of the coordinate fields take the form
 \begin{equation}
 T_{(\mu)}^{0i}=- \Pi_{(\mu)}\partial^i X_{(\mu)},\ \ \ \ \ \ 
T_{(\mu)}^{00}= 
 \frac{1}{2}\left(\Upsilon^{-1} \Pi_{(\mu)}^2 +\Upsilon \partial^i X_{(\mu)} \partial_i X_{(\mu)}\right) + V(X_{(\mu)}).
 \end{equation}
Using Eq.\,(\ref{XS}), the corresponding canonical momenta becomes $\Pi^{(\mu)} = \Upsilon \partial_0 X^{(\mu)} =\Upsilon (\delta^{(\mu)}_{0} +\partial_{0} \zeta^{(\mu)} (s) + \mathcal{O}(h^2))$. Therefore, we have
\begin{equation}\label{Momenta}
\Pi^{(0)} = \Upsilon(1 + \mathcal{O}(h)), \ \ \Pi^{(i)} \sim  \Upsilon \mathcal{O}(h). 
\end{equation}
We may then determine the stress-energy tensor associated with each scalar field $X^{(\mu)}$, with $\mu=0,1,2,3$, which we denote as $T_{(\mu)}^{\alpha\beta}$, with Greek letters labelling spacetime indices. In particular, the $\hat{T}_{(\mu)}^{0i}$ reads, up to our order of expansion: 
 \begin{equation}
T^{0i}_{(\mu)} = -\Pi_{(\mu)} \left(\delta^i_{(\mu)}+ \partial^{i} \zeta_{(\mu)}+  \mathcal{O}(h^2) \right).
\end{equation} 
The total stress-energy tensor of the reference field $\hat{X}^{(\mu)}$ involves the sum over the set of 4-scalars
$T^{0i} = \sum_{\mu} T^{0i}_{(\mu)} $:
 \begin{equation}\label{T0icondi}
T^{0i}  = \sum_{\mu}T^{0i}_{(\mu)}  = -\Pi^{(i)} -\sum_{\mu}  \Pi^{(\mu)}\left( \partial^{i} \zeta^{(\mu)} +  \mathcal{O}(h^2) \right) = -\Pi^{(i)} -\Upsilon \partial^i\zeta^{(0)} + \mathcal{O}(\Upsilon h^2),
\end{equation}
which means the leading terms are  linear in $\Pi^{(i)}$.

The calculation is less straightforward for the energy density $T^{00}$, but leads to an analogous expansion. For a given scalar field $X^{(\mu)}$, we have  
\begin{equation}
T^{00}_{(\mu)} =\frac{\Upsilon}{2} \left(\delta^0_{(\mu)}  +\partial^{0} \zeta_{(\mu)}\right)^2 +\frac{\Upsilon}{2}  \sum_i\left(\delta^i_{(\mu)}  +\partial^{i} \zeta_{(\mu)}  \right)^2  + V(X_{(\mu)}),
\end{equation} 
and hence
 \begin{equation}\begin{split}
T^{00}_{(0)} &= \Pi_{(0)} +\underbrace{  \frac{\Upsilon}{2} \left(\partial^{0} \zeta_{(0)} \right)^2 +\frac{\Upsilon}{2}\sum_i \left(\partial^i \zeta_{(0)} \right)^2 }_{\mathcal{O}(\Upsilon h^2)} + V(X_{(0)}) - \frac{\Upsilon}{2},\\
T^{00}_{(i)} &= \underbrace{ \frac{\Pi_{(i)}^2}{2 \Upsilon} }_{\mathcal{O}(\Upsilon h^2)} +\frac{\Upsilon}{2} \sum_j\left(\partial^{j} X_{(i)}  \right)^2  + V(X_{(i)}).
\end{split}
\end{equation} 
Summing the contributions of the four scalar fields, the leading-order of $T^{00}$ is 
\begin{equation}\label{T00condi}
T^{00}  =  \sum_{\mu} T^{00}_{(\mu)}=  \Pi_{(0)}  + f(X)+ \mathcal{O}(\Upsilon h^2),
\end{equation}
in which $f(X)$ only contains function of $X$ and their spatial derivatives. The specific functional form of $f(X)$ does not influence the result, it is only important that part commutes with $X$, which is automatically true. 
Note that, to our order of expansion, $T^{00}$ is positive definite. This can be easily seen by checking the leading order contribution to  $\Pi_{(0)}$, namely $\Pi_{(0)} = \Upsilon(1  +\partial_{0} \zeta^{(0)} (s)) $, which implies that the eigenvalues of $\Pi_0$ are positive because $\partial_{0} \zeta^{(0)}_{\alpha} (s) \ll 1.$

\section{Quantum states of the quantum reference fields}
\label{App:AdmiStates} 

Having derived the conditions under which classical reference fields are compatible with perturbative gravity,  we now specify how these conditions are imposed on the  quantum states of X-QRFs. Since quantum fields are operator-valued distributions, we formulate these conditions using smeared operators. We use a family of real, smooth test function $f_{\mathrm{s}}(\vec{u}) := f(\vec{u} - \vec{s})$ centred at $s^{\mu}$ to define the smeared field operator
\begin{equation}
\hat{X}^{(\mu)}_{f_s} (s^0) := \int d^3\vec{u} f_{\mathrm{s}}(\vec{u}) \hat{X}^{(\mu)} (s^0, \vec{u}), \ \ \hat{\Pi}^{(\mu)}_{f_s} (s^0)  := \int d^3\vec{u}  f_{\mathrm{s}}(\vec{u}) \hat{\Pi}^{(\mu)} (s^0, \vec{u} ),
\end{equation}
in which the smearing functions $f_{\mathrm{s}}$ have support of characteristic spatial size $\sigma$ centred around $s^{\mu}$, with normalisation $\int d^3\vec{u} f_{\mathrm{s}}(\vec{u})=1$. The smeared equal-time canonical commutation relation $[ \hat{X}^{(\mu)}_{f_s},\hat{\Pi}^{(\nu)}_{f_s}  ]= i \hbar \delta^{\mu\nu} \int d^3\vec{u} f_s(\vec{u})^2$
then gives the uncertainty relation 
\begin{equation}
\Delta \hat{X}^{(\mu)}_{f_s} \Delta \hat{\Pi}^{(\mu)}_{f_s}  \geq \frac{\hbar}{2} \int d^3\vec{u} f_s(\vec{u})^2.
\end{equation}
Consistently with Eq.(\ref{Momenta}), we require the momentum fluctuations of admissible X-QRF states to satisfy 
\begin{equation}\label{PIuncertainty}   
\Delta \hat{\Pi}^{(\mu)}_{f_s} = \sqrt{\langle (\hat{\Pi}^{(\mu)}_{f_s})^2 \rangle - \langle \hat{\Pi}^{(\mu)}_{f_s}\rangle^2  }  \leq  \Upsilon  \mathcal{O}(h).
\end{equation}
This requirement is independent of the usual argument based on gravitational backreaction, but is consistent with the expectation to bound matter momentum fluctuations in the weak-field regime. The uncertainty relation therefore gives
\begin{equation}\label{Xuncertainty}
\Delta \hat{X}^{(\mu)}_{f_s}  =\sqrt{\langle (\hat{X}^{(\mu)}_{f_s} )^2 \rangle - \langle \hat{X}^{(\mu)}_{f_s} \rangle^2  }   \ge \frac{\hbar}{2  \Upsilon  \mathcal{O}(h) }    \int d^3\vec{u} f_s(\vec{u})^2
\end{equation}
i.e. there is a lower bound on the uncertainty of the corresponding X-QRF. This means that the admissible X-QRF states cannot be arbitrarily sharply peaked on a definite field configuration.

Squeezed coherent states provide natural candidates for admissible state of X-QRFs.  Let us denote them as $\{ |\gamma^{\alpha}_{\mathrm{p}}  \rangle\}$, in which $\mathrm{p}:=(\mathrm{x,\pi})$ labels the expectation value of the field and momenta, and $\alpha$ labels  different width of the squeezed coherent states. 
An admissible state is such that  Eq.(\ref{PIuncertainty}, \ref{Xuncertainty}) are valid. In addition, the following conditions hold, given a fixed profile of test function $f$:
\begin{enumerate}
\item  The expectation values of the smeared field operator are smooth and locally invertible functions, analogous to classical coordinates: 
\begin{equation}\label{meanfield}
\langle \gamma^{\alpha}| \hat{X}^{(\mu)}_{f_s}  |\gamma^{\alpha}\rangle = \int d^3\vec{u} f_s(\vec{u})\  x^{(\mu)} (s^0,\vec{u}) := \mathrm{x}_f^{(\mu)}(s).
\end{equation}

\item  The quantum analogue of the condition in Eq.(\ref{backgroundcond}) holds
\begin{equation}
||\partial_{\sigma} \hat{\zeta}^{(\mu)}_A |\gamma^{\alpha}\rangle ||\sim\mathcal{O} ( h),
\end{equation}
in which $||...||$ is the norm of the state. $\hat{\zeta}^{(\mu)}_A (s) := \hat{X}^{(\mu)}_A(s) - s^\mu \hat{\mathbb{1}}_A$ is a displacement operator that indicates the difference between $\hat{X}^{(\mu)}_A(s)$ and an arbitrarily chosen background coordinate.  

\item  The quantum analogue of the condition in Eq.(\ref{Momenta}) is valid
\begin{equation}
|| (\hat{\Pi}^{(0)} - \Upsilon \hat{\mathbb{1}})|\gamma^{\alpha}\rangle || = \Upsilon \mathcal{O}(h) , \ \ || \hat{\Pi}^{(i)} |\gamma^{\alpha}\rangle ||\sim  \Upsilon \mathcal{O}(h). 
\end{equation}
\end{enumerate}
Since $\hat{T}^{0i}_A= - \frac{1}{2} \sum_{\mu=0}^3 \{\hat{\Pi}^{(\mu)}_A, \partial^i\hat{X}^{(\mu)}_A\} $
The classical conditions Eq.(\ref{T0icondi}, \ref{T00condi}) become operator matrix-element expansions  on the admissible XQRF states: 
\begin{equation}
\langle \gamma^{\alpha}_\mathrm{p}|(\hat{T}^{0i}_A)_{f_s} |\gamma^{\beta}_\mathrm{p'}  \rangle_A  = - \langle \gamma^{\alpha}_\mathrm{p}|(\hat{\Pi}^{i}_A)_{f_s}|\gamma^{\beta}_\mathrm{p'}  \rangle_A  - \Upsilon \langle \gamma^{\alpha}_\mathrm{p}|(\partial^i \hat{\zeta}_A^{(0)})_{f_s}|\gamma^{\beta}_\mathrm{p'}  \rangle_A   + \mathcal{O}(\Upsilon  h^2),   \ \ \forall \alpha, \beta, \mathrm{p}, \mathrm{p'}
\end{equation}
\begin{equation}
\langle \gamma^{\alpha}_\mathrm{p}|(\hat{T}^{00}_A )_{f_s}|\gamma^{\beta}_\mathrm{p'}  \rangle_A  =  \langle \gamma^{\alpha}_\mathrm{p}|(\hat{\Pi}^{(0)}_A)_{f_s}|\gamma^{\beta}_\mathrm{p'}  \rangle_A   +  \langle \gamma^{\alpha}_\mathrm{p}|F(\hat{X}^{(\mu)}_A)_{f_s}|\gamma^{\beta}_\mathrm{p'}  \rangle_A  + \mathcal{O}(\Upsilon h^2), \ \ \forall \alpha, \beta, \mathrm{p}, \mathrm{p'}
\end{equation}
Since the smearing profile $f_s$ is  arbitrarily chosen,  for notation simplicity, from now on we will also suppress the smeared field $f_s$, and write $\hat{X}^{(\mu)}_{f_s}(s^0)$ simply as $\hat{X}^{(\mu)}(s)$. Throughout the paper, all the operator equations involving perturbative expansions should be understood as the matrix element of the operator corresponding to the admissible state, i.e. $\langle \gamma^{\alpha}_\mathrm{p}|....|\gamma^{\beta}_\mathrm{p'}  \rangle_A $, unless specified otherwise.

To keep the time dependence explicit, both in the main text and in the  appendix, we work in the Heisenberg picture from now on. For the \QCF{}, the time evolution is governed by the Hamiltonian
\begin{equation}
\hat{H}_A^{(\mu)} (t) = \int d^3\vec{s} \  \hat{T}^{(\mu)}_{00}(t, \vec{s}) + \hat{h}^{ij} (t,\vec{s})\hat{T}^{(\mu)}_{ij} (t,\vec{s}).
\end{equation} 

The time evolution of the \QCF{}s is the following: 
\begin{equation}
\begin{split}
\hat{X}^{(\mu)}_A(s^0 + \Delta t, \vec{s}) & =  \overrightarrow{\mathrm{T}}  e^{\frac{i}{\hbar} \int_{s^0}^{s^0+\Delta t} \hat{H}_A^{(\mu)} (t) d t }\cdot \hat{X}^{(\mu)}_A(s^0, \vec{s})  \cdot  \overrightarrow{\mathrm{T}}  e^{ -\frac{i}{\hbar} \int_{s^0}^{s^0+\Delta t} \hat{H}_A^{(\mu)} (t) d t } \\
\end{split}
\end{equation}
In particular, for the \QCF{}  $\hat{X}^{(0)}_A$, we have at the leading order:
\begin{equation}
\hat{X}^{(0)}_A(s^0 + \Delta t, \vec{s})  = e^{\frac{i}{\hbar} \int d^3\vec{s}\ \hat{\Pi}^{(0)}(s^0, \vec{s}) \Delta t}\cdot \hat{X}^{(0)}_A(s^0, \vec{s})\cdot e^{-\frac{i}{\hbar} \int d^3\vec{s}\ \hat{\Pi}^{(0)}(s^0, \vec{s})\Delta  t} = \hat{X}^{(0)}_A(s^0, \vec{s}) +\Delta t \hat{\mathbb{1}}.
\end{equation}
Although a generic quantum state of the field $\hat{X}^{(0)}_A$ gives a superposition of quantum temporal coordinates, due to Eq.\eqref{backgroundcond},  all the eigenvalues of $\hat{X}^{(0)}_A(s^0)$ evolve at the same rate as the coordinate time at leading order, and deviate at higher orders. 

\section{Relational Diffeomorphism Invariant Observables}
\label{app:relob}
What counts as a physical observable is subtle in classical general
relativity. Diffeomorphism invariance, the gauge symmetry of the theory, eliminates the physical meaning of local spacetime points. The value of a field evaluated at certain manifold point is not diffeomorphism invariant, and in fact,  there are no non-trivial local physical observables in gravity. For the last few decades, there has been significant work on constructing physical observables in gravity ~\cite{Rovelli1991,Rovelli:2001bz, Dittrich:2005kc, Dittrich2007,Tambornino:2011vg, Brunetti:2013maa, Giddings:2025xym, Baldazzi:2021fye,Goeller:2022rsx}. One approach is to construct \emph{gravitationally dressed observables}, which are composite operators that couple bare matter excitations to gravitational degrees of freedom so that the whole operator is gauge invariant~\cite{Giddings:2025xym, Donnelly:2016rvo}. The idea can be traced back to Dirac’s gauge-invariant formulation of QED. Another related approach is to define \emph{relational observables} -- quantities specified relative to other dynamical systems, thus achieving gauge invariance without referring to coordinates or a spacetime background~\cite{Rovelli1991,Rovelli:2001bz, Dittrich:2005kc, Dittrich2007,Tambornino:2011vg, Brunetti:2013maa}. In classical gravity these two approaches can be related~\cite{Goeller:2022rsx}. 

Here we briefly introduce relational observables \cite{Brunetti:2013maa, Giddings:2025xym, Baldazzi:2021fye} in the language of the reference fields. 
 For a classical reference field $X^{(\mu)}_A (s)$, we define the inverse map $\chi(\mathrm{y}^{(\mu)})$ through
\begin{equation} 
X^{(\mu)}_A(\chi_A(\mathrm{y}^{(\mu)})) = \mathrm{y}^{(\mu)}.
\end{equation}
Here $\{\mathrm{y}^{(\mu)}\}$ is  the set of physical value of the reference field. The inverse map $\chi(\mathrm{y}^{(\mu)})$ therefore identifies the manifold point at which the reference fields take the value $\mathrm{y}^{(\mu)}$.
Given a scalar field $O(s)$, the corresponding relational observable is constructed as 
\begin{equation}
\tilde{O}(\mathrm{y}^{(\mu)}) := O (\chi_A(\mathrm{y}^{(\mu)})).
\end{equation}
This quantity means: evaluate $O$ at the point where the reference field $X^{\mu}_A$ take the value $\mathrm{y}^{(\mu)}$. It is also called \emph{complete observable} in \cite{Rovelli:2001bz}.
It can be easily checked that this observable is invariant under diffeomorphism.  Consider an infinitesimal diffeomorphisms in the direction of $\xi^{\mu}$. From the identity
\begin{equation}
\delta_{\xi} X_A^{(\nu)} \left(\chi_A(\mathrm{y}^{(\nu)})\right) = \xi^{\mu}\partial_{\mu}  X_A^{(\nu)} \left(\chi_A(\mathrm{y}^{(\nu)})\right) +  \left[\partial_{\mu}X_A^{(\nu)} \left(\chi_A(\mathrm{y}^{(\nu)})\right) \right] \cdot \delta_{\xi}\left(\chi_A(\mathrm{y}^{(\mu)})\right) \equiv 0,
\end{equation}
one obtains
\begin{equation}
\delta_{\xi}\left(\chi_A(\mathrm{y}^{(\mu)})\right)  = - \xi^\mu,
\end{equation}
where we have omitted the argument of $\xi_\mu $, i.e.\,$ \xi_\mu(\chi_A(\mathrm{y}^{(\mu)}))$. Therefore, we have
\begin{equation}
\delta_{\xi} O \left(\chi_A(\mathrm{y}^{(\mu)})\right) = (\xi^{\mu}\partial_{\mu}  O)\left(\chi_A(\mathrm{y}^{(\mu)})\right) + \partial_\mu  O\left(\chi_A(\mathrm{y}^{(\mu)})\right) \cdot \delta_{\xi} \left(\chi_A(\mathrm{y}^{(\mu)})\right) = 0
\end{equation}
In summary, they are defined by evaluating a field where another physical field takes a specified value. This construction allows one to remove the need to refer to unphysical background coordinates and is automatically gauge invariant.

The same idea  can be extended to tensors.  The only difference is that one must also take care of the tensor indices transformation. For metric field, the relational Dirac observable is
\begin{equation}\label{eq:ClassRelMetric}
\tilde{g}_{\alpha \beta}(\mathrm{y}^{(\mu)}) := J^{\sigma}_\alpha J^{\rho}_\beta g_{\sigma \rho} (\chi_A(\mathrm{y}^{(\mu)})),
\end{equation}
where 
\begin{equation}
J^{\mu}_\alpha: = \frac{\partial \chi_A (\mathrm{y}^{(\mu)})}{\partial \mathrm{y}^{(\alpha)}},
\end{equation}
is the Jacobian of the inverse map. The indices $\alpha, \beta$ labels the physical reference field, 
and $y^{(\mu)}$ is the physical value of the reference field $X^{(\mu)}_A$. With the same approach as above, one can check that 
\begin{equation}
\delta_{\xi}\tilde{g}_{\alpha \beta}(\mathrm{y}^{(\mu)})  =0.
\end{equation}
 More generally, for any tensor field, the corresponding relational observable is obtained by evaluating the tensor at $\chi_A(\mathrm{y}^{(\mu)})$ and transforming the tensor indices into the indices of physical coordinate system defined by the reference fields.

We now extend the inverse map $\chi_A$  to quantum reference fields at the fixed spatial resolution $\sigma$ introduced in Appendix C. 
We use $\hat{\mathrm{Y}}^{(\mu)}_s$ to denote a set of non-dynamical operators - the \QCF{} operator $\hat{X}^{(\mu)}_A(s)$ evaluated at a fixed $s^{\mu}$, with smeared version $\hat{\mathrm{Y}}^{(\mu)}_{f_s}$. Due to the assumption in Eq.(\ref{meanfield}), we have
\begin{equation}
\langle \gamma^{\alpha}| \hat{Y}^{(\mu)}_{f_s}  |\gamma^{\alpha}\rangle = \int d^3\vec{s} f_s(\vec{s})\  x^{(\mu)} (s^0,\vec{s}) = \mathrm{x}_f^{(\mu)}(s).
\end{equation}
The inverse map $\chi_A$ takes as an input $\hat{\mathrm{Y}}^{(\mu)}_s$ and returns an identity operator that is proportional to the background coordinate $\hat{\mathbb{1}} s^{\mu}$:
\begin{equation}\label{eq:chidef}
\chi_A(\hat{\mathrm{Y}}^{(\mu)}_{f_s})  = s^{\mu} \hat{\mathbb{1}} 
\end{equation}
We can easily see why the inverse map exist:  if two points $s_1$ and  $s_2$ give the same operator $\hat{\mathrm{Y}}_{f_s}$, then we have $\langle \gamma^{\alpha}| \hat{\mathrm{Y}}^{(\mu)}_{f_{s1}} | \gamma^{\alpha}\rangle = \langle \gamma^{\alpha}| \hat{\mathrm{Y}}^{(\mu)}_{f_{s2}} | \gamma^{\alpha}\rangle =  \mathrm{x}_{f_s}^{(\mu)}(s)$. By the invertability of $\mathrm{x}_{f_s}^{(\mu)}(s)$, this means $s_1 = s_2 = s$. Therefore, having the mean field of admissible states that are classical coordinates establish the existence of $\chi_A$.

The quantum relational observable for the metric field generalizes the classical expression in Eq.\,\eqref{eq:ClassRelMetric} to
\begin{equation}
\hat{J}^{\mu}_{\alpha} \hat{J}^{\nu}_{\beta} g_{\mu \nu} \left(\chi_A(\hat{\mathrm{Y}}_{f_s}^{(\sigma)}) \right),
\end{equation}
where we have defined the Jacobian operator by analogy to the classical one, namely $\hat{J}^{\mu}_{\alpha}  : =  \frac{\partial \chi_A(\hat{\mathrm{Y}}_{f_s}^{(\mu)})}{\partial \mathrm{\hat{\mathrm{Y}}}_{f_s}^{\alpha}}$.
Similarly, for a general $(m,n)$ tensor field $\hat{F}_{\mu..}^{\nu..}(s)$, the relational field operator can be constructed accordingly as
\begin{equation}
 \hat{J}^{\mu}_{\alpha}  ...\hat{J}_{\nu}^{\beta} ...\hat{F}_{\mu..}^{\nu..} (\chi_A(\hat{\mathrm{Y}}_{f_s}^{(\sigma)})).
\end{equation}
In the other sections, we see how these observables are represented in a \QCF{} perspective, and how they transform between perspectives. 

\section{The Constraint Trivilization Map}
\label{app:Tri}
In this section, we derive the action of constraint-trivilization map, defined in Eq.\,\eqref{TriMap} in the main text, on the constraints.
In the Heisenberg picture, the trivialisation map to the \QCF{} $A$ is defined as
\begin{equation}
 \hat{\mathcal{T}}_{A} (s^0) = e^{-  \frac{i}{\hbar}  \int d^3s  \sum_{\mu=0}^{3} \hat{\zeta}^A_{(\mu)}(s) \left(\hat{T}^{0\mu}_B(s)+\hat{T}^{0\mu}_S(s) +2\partial_{\nu} \hat{\pi}_G^{\nu \mu} (s) + \delta_0^\mu  \frac{1}{\kappa}\Delta \hat{h}^T(s)\right)}.
\end{equation} 
where we use $s:= s^{\mu}$ for brevity, $\hat{\zeta}^{(\mu)}_A (s):=\hat{X}^{(\mu)}_A(s)  - \hat{\mathbb{1}} s^{(\mu)}$, and $\hat{\zeta}^A_{(\mu)}(s):= \eta_{\mu \nu} \hat{\zeta}_A^{(\nu)}(s)$.

The above map has the structure of $ \hat{\mathcal{T}}_{A}  := e^{-\frac{i}{\hbar}\hat{K}(s^0)}.$
Acting on the constraints $\hat{\mathcal{C}}^{\mu} (s)$ at $s^0$,  we have 
\begin{equation}\label{BCH}
	 \hat{\mathcal{T}}_{A}  \cdot \hat{\mathcal{C}}^{\mu} (s) 	\cdot  \hat{\mathcal{T}}_{A} ^{\dagger}  := e^{-\frac{i}{\hbar}\hat{K}(s^0)}\hat{\mathcal{C}}^{\mu}(s^0, \vec{s})  e^{\frac{i}{\hbar}\hat{K}(s^0)} = \hat{\mathcal{C}}^{\mu}- \frac{i}{\hbar}\left[ \hat{K},\hat{\mathcal{C}}^{\mu}\right]+ \frac{i^2}{2 \hbar^2}  \left[\hat{K}, \left[ \hat{K}, \hat{\mathcal{C}}^{\mu}\right]\right]+ ...
\end{equation}
In Appendix~\ref{app:CoordFields}, we have derived the expansion of the stress-energy tensor for \QCF{}s (see in particular Eq.\,\eqref{T0icondi}). Specifically, for a coordinate field $\hat{X}_A$, we have
 \begin{equation}
\hat{T}^{0i}_{A}  = \sum_{\mu}\hat{T}^{0i}_{A, (\mu)}  = -\hat{\Pi}^{(i)}_{A} -\sum_{\mu}  \hat{\Pi}^{(\mu)}_{A} \partial^{i} \hat{\zeta}^{(\mu)}_{A} + \sum_{\mu}\hat{\Pi}^{(\mu)}_{A}  \mathcal{O}(h^2),
\end{equation} 
in which $\hat{\Pi}_{(\mu)}^{A}$ are the canonical momenta of $\hat{X}_{A}^{(\mu)}$, $\ \mu = 0,1,2,3$,
and they satisfy the commutation relation
\begin{equation}
\left[\hat{X}^{J}_{(\nu)} (s^0, \vec{s}), \hat{\Pi}_{J}^{(\mu)}(s^0, \vec{s}')\right] = i \hbar \delta^\mu_\nu \delta^{(3)}\left(\vec{s} - \vec{s}'\right).
\end{equation}
Crucially, for the \QCF{}s satisfying Eq.\,\eqref{backgroundcond}, the operator $\hat{\zeta}^{(\mu)}_A (s)=\hat{X}^{(\mu)}_A(s)  - \hat{\mathbb{1}} s^{(\mu)}$  satisfies the same commutation relations as $X_A^{(\mu)}$, because they only differ by a term proportional to the identity operator, namely 
\begin{equation}
\left[\hat{\zeta}_{A}^{(\nu)} (s^0, \vec{s}), \hat{\Pi}^{A}_{(\mu)}(s^0, \vec{s}')\right] = \left[\hat{X}_{A}^{(\nu)} (s^0, \vec{s}) - \hat{\mathbb{1}} s^{(\nu)}, \hat{\Pi}^{A}_{(\mu)}(s^0, \vec{s}')\right] = i \hbar \delta^\mu_\nu \delta\left(\vec{s} - \vec{s}'\right).
\end{equation}
Now let us come back to Eq.\,\eqref{BCH}.
For the first commutator, we have at the leading order 
\begin{equation}\label{eq:KCicommutator}
\begin{split}
\left[ \hat{K}(s^0),\hat{\mathcal{C}}^i(s^0, \vec{s}')\right] &= \left[   \int d^3s  \sum_{\mu} \hat{\zeta}^A_{(\mu)}(\vec{s}) \left(\hat{T}^{0\mu}_B(\vec{s})+\hat{T}^{0\mu}_S(\vec{s}) +2\partial_j \hat{\pi}_G^{j\mu} (\vec{s})\right) , \hat{T}^{0i}_A(\vec{s}') +\hat{T}^{0i}_B(\vec{s}')+\hat{T}^{0i}_S(\vec{s}') +2\partial_j \hat{\pi}_G^{ij} (\vec{s}') \right]_{s^0}\\
&= - \left[ \int d^3s   \hat{\zeta}^A_{(\mu)}(\vec{s}) \left(\hat{T}^{0\mu}_B(\vec{s})+\hat{T}^{0\mu}_S(\vec{s}) +2\partial_j \hat{\pi}_G^{\mu j} (\vec{s})\right),\hat{\Pi}^{(i)}_A(\vec{s}') (1 + \mathcal{O} (h)) +\hat{T}^{0i}_S(\vec{s}')\right]_{s^0} \\
&= - i \hbar \left(\hat{T}^{0i}_B(s^0, \vec{s}')+\hat{T}^{0i}_S(s^0,\vec{s}') +2\partial_j \hat{\pi}_G^{ij} (s^0,\vec{s}')\right) (1 + \mathcal{O} (h)) - \sum_{\mu \neq i}  \int d^3s   \hat{\zeta}^A_{(\mu)}(\vec{s}) \left[\hat{T}^{0\mu}_S(\vec{s}), \hat{T}^{0i}_S(\vec{s}')\right]_{s^0} 
\end{split}
\end{equation}
in which the last commutator is given by the Schwinger terms~\cite{Schwinger:1959xd, Deser:1967zzf}
\begin{equation}
[T^{0l}_S(\vec{s}), T^{0i}_S(\vec{s}')] = - i \hbar \left(T^{0i}_S(\vec{s}) \partial_l \delta^{(3)} (\vec{s}-\vec{s}') - T^{0l}_S(\vec{s}) \partial_i \delta^{(3)} (\vec{s}-\vec{s}') \right).
\end{equation}
Plugging the previous expression into the last term of Eq.\,\eqref{eq:KCicommutator}, we obtain
\begin{equation}
 \sum_{\mu \neq i} \left[ \int d^3s   \hat{\zeta}^A_{(\mu)}(\vec{s}) \hat{T}^{0\mu}_S(\vec{s}), \hat{T}^{0i}_S(\vec{s}')\right]_{s^0} = (\partial_j\hat{\zeta}_A^j(\vec{s}')) \cdot \hat{T}^{0i}_S(\vec{s}') + i\hbar\hat{\zeta}^{(j)}_A(\vec{s}') \partial_{j} \hat{T}^{0i}_S (\vec{s}') +i\hbar\hat{\zeta}^{(0)}_A(\vec{s}') \partial_{j} \hat{T}^{ij}_S (\vec{s}'). 
\end{equation}
We need the previous quantity to be of order $\mathcal{O}(h^2)$, so that it can be neglected within our approximation. The first term is already of order $\mathcal{O}(h^2)$, since $\hat{T}^{0i}_S(\vec{s}')=\mathcal{O}(h)$. For the remaining terms, we require one of the following two conditions to be met: 
\begin{enumerate}
\item The spatial derivative of the source is small $\partial_{j}\hat{T}^{0i} (\vec{s}') \sim \mathcal{O}(h^2)$ 
\item The value of quantum reference field is very close to the background coordinate chart $s$: $\hat{\zeta}^{(j)} = \hat{X}^{(j)} - s^j \hat{\mathbb{1}} \sim \mathcal{O}(h)$. 
\end{enumerate}
If one of the above conditions is satisfied, the second commutator $ \left[\hat{K}, \left[ \hat{K}, \hat{\mathcal{C}}^i(\vec{s})\right]\right]$ vanishes within our order of approximation. We obtain, at any $s^0$
\begin{equation}
	 \hat{\mathcal{T}}_{A}  \cdot \hat{\mathcal{C}}^{i} (s) 	\cdot  \hat{\mathcal{T}}_{A} ^{\dagger}  =  \hat{T}^{0i}_A(s) \left(1 - \mathcal{O}(h)\right).
\end{equation}
A similar derivation holds for the scalar constraint. We find 
\begin{equation}
\begin{split}
\left[ \hat{K}(s^0),\hat{\mathcal{C}}(s^0, \vec{s}')\right] &= \left[  \int d^3s    \hat{\zeta}^A_{(0)} (\vec{s})\left( \hat{T}^{00}_B(\vec{s})+ \hat{T}^{00}_S(\vec{s}) + \frac{1}{\kappa}\Delta \hat{h}^T (\vec{s})\right), \hat{T}^{00}_A(\vec{s}')\right]_{s^0}\\
&=  \left[ \int d^3s  \hat{\zeta}^A_{(0)} (\vec{s})\left(\hat{T}^{00}_B(\vec{s})+ \hat{T}^{00}_S(\vec{s}) + \frac{1}{\kappa}\Delta \hat{h}^T (\vec{s})\right),\hat{\Pi}_{(0)}^A(\vec{s}') +f(X_{A})+ \mathcal{O}(\Upsilon h^2)\right]_{s^0} \\
&= - i \hbar \left(\hat{T}^{00}_B(s^0,\vec{s}')+\hat{T}^{00}_S(s^0,\vec{s}') + \frac{1}{\kappa}\Delta \hat{h}^T (s^0,\vec{s}')\right) + \mathcal{O}(\Upsilon h^2).
\end{split}
\end{equation}
where the minus sign in the final line comes from $\hat{\zeta}^A_{(0)}(s)= - \hat{\zeta}_A^{(0)}(s)$, due to the temporal signature of Minkowski metric. Therefore we arrive at 
\begin{equation}
	\hat{\mathcal{T}}_{A} \cdot  \hat{\mathcal{C}} (s) \cdot	\hat{\mathcal{T}}^{\dagger}_{A}=  \hat{T} ^{00}_{A}(s) + \mathcal{O}(h^2).
\end{equation}
Notice that one could in principle change the trivialization map to avoid imposing additional restrictions on the stress energy tensor and on the reference fields. However, the construction would become significantly more complicated, and we have not attempted this route.

Acting on the \QCF{}s, the trivilization maps give us
\begin{equation}
\hat{\mathcal{T}}_{A}  (s^0)\cdot \hat{X}^{(\mu)}_B (s^0, \vec{s}) \cdot \hat{\mathcal{T}}^{\dagger}_{A}  (s^0)=  \hat{X}^{(\mu)}_{B|A}(s^0, \vec{s}) +  \hat{\zeta}^{(\mu)}_{A|A}(s^0, \vec{s}),
\end{equation}
\begin{equation}
\hat{\mathcal{T}}_{A} (s^0) \cdot \hat{X}^{(\mu)}_A  (s^0, \vec{s}) \cdot \hat{\mathcal{T}}^{\dagger}_{A} (s^0) =  \hat{X}^{(\mu)}_{A|A}(s^0, \vec{s}).
\end{equation}
In the second equation, we have used that $\hat{\mathcal{T}}_{A}$ commutes with  $\hat{X}^{(\mu)}_A$ and only changes the Hilbert space labelling to $ \hat{X}^{(\mu)}_{A|A}$. This is the formal expression of $\hat{X}^{(\mu)}_A$ in the perspective of itself, which serves as the ``quantum coordinates'' for the observables in the reduced perspective.

\section{Derivation of the Perspectival Observables}
\label{app:TMetric}
In this section, we derive the perspectival observables through evaluating the action of the constraint-trivialisation map on the diffeomorphism invariant relational operators. In general, we define the trivialisation map as the operator that takes an element of the perspective neutral Hilbert space, or physical Hilbert space $\mathcal{H}^{PN}_{ABSG}$ to an element of the isomorphic Hilbert space $\mathcal{H}_{ABSG|A}$. Importantly, the trivialisation map changes the partition of the total Hilbert space, and hence the definition of subsystems~\cite{Vanrietvelde:2018pgb, hoehn2021quantum}.

The condition on the \QCF{}s in Eq.\,(\ref{backgroundcond1}) leads to the expansion $ \hat{X}^{(\mu)}_A (s) = \hat{\mathbb{1}}_A s^{\mu}  + \hat{\zeta}_A^{(\mu)} (s)$, in which $ \partial_\nu \hat{\zeta}_A^{(\mu)} \sim \mathcal{O}(h)$. Therefore, the inverse map defined in Eq.\,\eqref{eq:chidef} can be explicitly written as
\begin{equation}\label{app:inverse}
\chi_A^{(\mu)} (\hat{\mathrm{Y}}_s) = \hat{\mathrm{Y}}_s^{(\mu)} - \hat{\zeta}^{(\mu)}_A(\mathrm{Y_s}) + \zeta_A^{(\alpha)}(\hat{\mathrm{Y}}_s) \partial_\alpha \zeta_A^{(\mu)}(\hat{\mathrm{Y}}_s)+\mathcal{O}(h^2)
\end{equation}
Notice that the index in the round brackets $(\mu)$ labels the four scalar fields. From now on, we sometimes omit the superscript ${(\mu)}$ in $\hat{\mathrm{Y}}^{(\mu)}_s$, and $s^0$ in $\hat{\mathcal{T}}_A (s^0)$ when no confusion arises.

For a diffeomorphism-invariant relational scalar field operator $\hat{\psi}_S(\chi_A(\hat{\mathrm{Y}}_s))$ in the perspective neutral structure, the trivilisation map gives us, in the perspective of the \QCF{} $A$, 
\begin{equation}\label{eq:scalar}
\begin{split}
\hat{\mathcal{T}}_A  \hat{\psi}_S(\chi_A(\hat{\mathrm{Y}}_s)) \hat{\mathcal{T}}^{\dagger}_A 
 &=   e^{-  \frac{i}{\hbar}  \int d^3s  \hat{\zeta}^A_{(\mu)}(s^0,\vec{s}) \hat{T}^{0\mu}_S(s^0,\vec{s})}  \cdot \hat{\psi}_S(\chi_A(\hat{\mathrm{Y}}_s))  \cdot  e^{ \frac{i}{\hbar}  \int d^3s  \hat{\zeta}^A_{(\mu)}(s^0,\vec{s}) \hat{T}^{0\mu}_S(s^0,\vec{s})}\\
 &=   \hat{\psi}_{S|A}\left(\chi^{(\mu)}_A(\hat{\mathrm{Y}}_s) + \hat{\zeta}_A^{(\mu)} (\chi_A(\hat{\mathrm{Y}}_s))\right) \\
  &=  \hat{\psi}_{S|A}\left(  \hat{\mathrm{Y}}_s^{(\mu)} - \hat{\zeta}^{(\mu)}_A(\mathrm{Y_s}) + \zeta_A^{(J)}(\hat{\mathrm{Y}}_s) \partial_J \zeta_A^{(\mu)}(\hat{\mathrm{Y}}_s) + \hat{\zeta}_A^{(\mu)} (\hat{\mathrm{Y}}_s  ) - \zeta_A^{(J)}(\hat{\mathrm{Y}}_s) \partial_J \zeta_A^{(\mu)}(\hat{\mathrm{Y}}_s)  + \mathcal{O}(h^2)\right) \\
        &=\hat{\psi}_{S|A} ( \hat{X}^{(\mu)}_{A|A}(s)). \\
\end{split}
\end{equation}
The above calculation gives us explicitly the scalar field $\psi_S$ expressed  in the coordinates of the \QCF{} $A$. The Hilbert space labelling  $|A$ denotes that the operators are written in the perspective of \QCF{} $A$.  In deriving the second line of Eq.\,\eqref{eq:scalar}, we have Taylor expanded the unitary operator, and then  used the infinitesimal action of the trivialisation map on the scalar field, namely:
\begin{equation}
i \big[\int d\vec{s}\ \hat{T}^S_{0\mu}(\vec{s}) \zeta^{\mu}(\vec{s}),  \hat{\psi}_S(\vec{x})\big]  = \zeta^{\mu}\partial_{\mu}\hat{\psi}_S(\vec{x}) =\mathcal{L}_{\zeta} \hat{\psi}_S(\vec{x}).
\end{equation}

Since \QCF{}s are a set of 4-scalar fields, from Eq.\,\eqref{eq:scalar} we immediately have the expression of coordinate field B in the description of coordinate field A:
\begin{equation}
\begin{split}
\hat{\mathcal{T}}_A  \hat{X}^{(\nu)}_B(\chi_A(\hat{Y}_s)) \hat{\mathcal{T}}^{\dagger}_A  &= \hat{X}^{(\nu)}_{B|A}(\hat{X}_{A|A}(s)).
\end{split}
\end{equation}
It is worth noting that observables, defined in the perspective neutral structure, of the type of
$\hat{X}_B^{(\mu)} (s) -\hat{X}_A^{(\mu)} (s) $ are in general not diffeomorphism invariant, therefore not the physical  observable in the gravitational context, as can be seen by checking
\begin{equation}
\mathcal{L}_{\xi} \left(\hat{X}_B^{(\mu)} (s) -\hat{X}_A^{(\mu)} (s) \right) = \xi^{\nu} \partial_{\nu} \left(\hat{X}_B^{(\mu)} (s) -\hat{X}_A^{(\mu)} (s) \right) =  \xi^{\nu} \partial_{\nu} \left(\hat{\zeta}_B^{(\mu)} (s) -\hat{\zeta}_A^{(\mu)} (s) \right) \neq 0. 
\end{equation}
This constitutes a difference with the particle case, where  the relative coordinates defined with respect to translation QRFs coincide with relational Dirac observables~\cite{Vanrietvelde:2018pgb}. 
Here the physical observable of the \QCF{} instead  is $\hat{X}_B(\chi_A(\hat{\mathrm{Y}}))$ in the PN space, and reduces to $\hat{X}_{B|A}(\hat{X}_{A|A})$ in $A$'s perspective.

We now evaluate the action of $\hat{\mathcal{T}}_A $ on the  relational metric field operator. 
First of all, it is useful to expand  the inverse Jacobian explicitly:
\begin{equation}
\hat{J}^{\mu}_{\alpha} =  \hat{\mathbb{1}}  \delta_{\alpha}^{\mu} - \partial_{\alpha} \hat{\zeta}^{(\mu)}_A(\hat{\mathrm{Y}}_s)  + \partial_{\alpha}\left(\hat{\zeta}_A^\beta(\hat{\mathrm{Y}}_s) \partial_\beta \hat{\zeta}_A^{(\mu)}(\hat{\mathrm{Y}}_s)  \right) .
\end{equation}
This allows us to expand the diffeomorphism invariant  relational metric explicitly as
\begin{equation}
\begin{split}
 \hat{J}^{\mu}_{\alpha}  \hat{J}^{\nu}_{\beta} \hat{g}_{\mu \nu} (\chi_A(\hat{\mathrm{Y}}_s)) & =  \hat{J}^{\mu}_{\alpha}  \hat{J}^{\nu}_{\beta}   \left( \eta_{\mu \nu}+ \hat{h}_{\mu \nu}(\chi_A(\hat{\mathrm{Y}}_s)) \right) \\
 = & \ \hat{\mathbb{1}}  \eta_{\alpha\beta} + \hat{h}_{\alpha\beta}(\chi_A(\hat{\mathrm{Y}}_s))  -\partial_{\alpha} \hat{\zeta}^A_{(\beta)} (\chi_A(\hat{\mathrm{Y}}_s))  -\partial_{\beta} \hat{\zeta}^A_{(\alpha)} (\chi_A(\hat{\mathrm{Y}}_s)).
 \end{split}
\end{equation}
Note that the Minkowski metric $\eta_{\alpha\beta}$ does not depend on any spacetime point or local field, and should be understood as multiplied by the identity operator. 
Upon quantization, $\eta_{\alpha\beta}$ remains classical background and $h_{\alpha\beta}$ is promoted to operator.  We now apply   the constraint-trivialization map to obtain the relational metric operator in the perspective of $A$. Using  the canonical commutator of the gravitational field $ \left[h_{0\mu} (\vec{x}), \pi^{0\nu} (\vec{x}') \right] =i \hbar  \delta^{\mu}_{\nu}  \delta^3(\vec{x}-\vec{x}')$ , we have 
\begin{equation}\label{trivilisationG}
\begin{split}
\hat{\mathcal{T}}_{A}  \hat{J}^{\mu}_{\alpha}  \hat{J}^{\nu}_{\beta}  \hat{g}_{\mu \nu} (\chi_A(\hat{\mathrm{Y}}_s)) \hat{\mathcal{T}}_{A}^{\dagger} &=\hat{J}^{\mu}_{\alpha} \hat{J}^{\nu}_{\beta}  \hat{\mathcal{T}}_{A} \hat{g}_{\mu \nu} (\chi_A(\hat{\mathrm{Y}}_s))\hat{\mathcal{T}}_{A}^{\dagger} 
 \\
&= \hat{J}^{\mu}_{\alpha}  \hat{J}^{\nu}_{\beta} \left(\hat{\mathbb{1}} \eta_{\mu \nu} + \hat{h}_{\mu \nu}(\chi_A(\hat{\mathrm{Y}}_s))  -\frac{2i}{\hbar}\int d^3s'  \sum_l \hat{\zeta}^A_{(l)}(\vec{s}') \partial_{\sigma} [\hat{\pi}_G^{l\sigma} (\vec{s}'),\hat{h}_{\mu \nu}(\chi_A(\hat{\mathrm{Y}}_s))] \right) \\
&= \hat{J}^{\mu}_{\alpha}  \hat{J}^{\nu}_{\beta} \left(\hat{\mathbb{1}} \eta_{\mu \nu} + \hat{h}_{\mu \nu}(\chi_A(\hat{\mathrm{Y}}_s)) + \partial_{\mu}\hat{\zeta}_{\nu} (\chi_A(\hat{\mathrm{Y}}_s))+  \partial_{\nu}\hat{\zeta}_{\mu} (\chi_A(\hat{\mathrm{Y}}_s)) \right)\\
&=\hat{\mathbb{1}} \eta_{\alpha\beta}    + \hat{h}_{\alpha\beta}(\chi_A(\hat{\mathrm{Y}}_s))   -\partial_{\alpha} \hat{\zeta}^A_{(\beta)} (\chi_A(\hat{\mathrm{Y}}_s))  -\partial_{\beta} \hat{\zeta}^A_{(\alpha)} (\chi_A(\hat{\mathrm{Y}}_s)) + \partial_{\alpha} \hat{\zeta}^A_{(\beta)}(\chi_A(\hat{\mathrm{Y}}_s))  + \partial_{\beta} \hat{\zeta}^A_{(\alpha)}(\chi_A(\hat{\mathrm{Y}}_s)) \\
&= \hat{\mathbb{1}} \eta_{\alpha\beta}   + \hat{h}_{\alpha\beta}(\chi_A(\hat{\mathrm{Y}}_s))
\end{split}
\end{equation}
where the $\alpha,\beta$ indices mean that the tensor is expressed in the coordinate field $\hat{X}_A$. We denote the perspectival metric by a more compact notation as
\begin{equation}
\hat{g}_{\alpha\beta}^{A} :=\hat{\mathbb{1}} \eta_{\alpha\beta}   + \hat{h}_{\alpha\beta}(\chi_A(\hat{\mathrm{Y}}_s)).
\end{equation}
Note that in the regime when all the eigenfunctions of  $\hat{X}^{(\mu)}_A$ are very close to the initial background coordinates, the field operator can be approximated by $\hat{X}^{(\mu)}_A - \hat{\mathbb{1}}s^{\mu} = \hat{\zeta}_A^{(\mu)} \sim \mathcal{O}(h)$. This is also the Condition 2 in Appendix~\ref{app:Tri}. In this case, the above expression can be further expanded in the similar form as Eq.\eqref{eq:scalar}:
\begin{equation}\label{smallappro}
\begin{split}
\hat{h}_{\alpha\beta}(\chi_A(\hat{\mathrm{Y}}_s)) &= \hat{h}_{\alpha\beta}\left( \hat{\mathrm{Y}}_s^{(\mu)} - \hat{\zeta}^{(\mu)}_A(\mathrm{Y_s}) + \zeta_A^{\gamma}(\hat{\mathrm{Y}}_s) \partial_{\gamma} \zeta_A^{(\mu)}(\hat{\mathrm{Y}}_s) \right)  \\
&= \hat{h}_{\alpha\beta}(\hat{\mathrm{Y}}_s) - \hat{\zeta}_A^{\gamma}\partial_{\gamma}\hat{h}_{\alpha\beta}(\hat{\mathrm{Y}}_s) + \mathcal{O}(\hat{h}^3) \\
&= \hat{h}_{\alpha\beta}(\hat{\mathrm{Y}}_s) + \mathcal{O}(\hat{h}^2)\\
&= \hat{h}_{\alpha\beta}(\hat{X}_{A|A}(s)) + \mathcal{O}(\hat{h}^2).\\
\end{split}
\end{equation}
From the reduction of the relational observables, we can see that the \QCF{}s provide  physical ``coordinates" and relational localization for the systems $\hat{\psi}_S, \hat{h}_{\mu \nu}$, and so on, in the perspective of each reference field.

\section{The Transformations Between Quantum Coordinate Fields}
\label{app:Trans}

In this section, we derive the perspectival transformation between \QCF{}s. Let us consider a scalar field observable to illustrate the idea. Since the constraint trivilization \eqref{TriMap} is invertible, 
its inverse maps the perspectival observables $\psi (\hat{X}_B)$ in the description relative to one \QCF{} to the corresponding Dirac observable  in the PN space: $\hat{\mathcal{T}}_B \psi (\hat{X}_B) \hat{\mathcal{T}}^{\dagger}_B= \psi(\chi_B(\hat{\mathrm{Y}}_s))$.  
Transforming to the description relative to another coordinate field A is to relate $\psi(\chi_B(\hat{\mathrm{Y}}_s))$ and $\psi(\chi_A(\hat{\mathrm{Y}}_s))$. Suppose that the 
inverse maps $\chi_A(\hat{\mathrm{Y}}_s^A)$ and $\chi_B(\hat{\mathrm{Y}}_s^B)$ identify the same point on the underlying manifold,  parametrized by the background chart  $s^{\mu}$, namely
\begin{equation}\label{samepoint}
\chi_A({\hat{\mathrm{Y}}^A_s}) = \chi_B({\hat{\mathrm{Y}}^B_s}) = s^{\mu} \hat{\mathbb{1}}_{A,B}.
\end{equation}
Then both $\psi(\chi_A(\hat{\mathrm{Y}}_s))$  and $\psi(\chi_B(\hat{\mathrm{Y}}_s))$ encode the  information of the system at the same location. 
The transformation between the corresponding reduced descriptions is expressed as follows: 
\begin{equation}\label{passiveQCF}
\hat{S}_{B\rightarrow A} :=   \hat{\mathcal{T}}_{A}  \cdot  \hat{\mathcal{T}}^{\dagger}_{B}    = e^{-\frac{i}{\hbar} \int d^3s  \hat{\zeta}^A_{(\mu)}(\vec{s}) \hat{T}^{0\mu}_B(\vec{s})  } \cdot  e^{-\frac{i}{\hbar}  \int d^3s   \hat{\xi}^{A|B}_{(\mu)}(s) \left(\hat{T}^{0\mu}_S(s) +2\partial_{\nu} \hat{\pi}_G^{\nu \mu} (s) + \delta_0^\mu  \frac{1}{\kappa}\Delta \hat{h}^T(s)\right)}  \cdot  e^{\frac{i}{\hbar} \int d^3s  \hat{\zeta}^B_{(\mu)}(\vec{s}) \hat{T}^{0\mu}_A(\vec{s})  }. 
\end{equation}
where $\hat{\xi}^{A|B}_{(\mu)}(s):= \hat{\zeta}^{A}_{(\mu)}(s) -\hat{\zeta}^{B}_{(\mu)}(s)$.
The above expression is written in terms of operators in the perspective neutral.  However, since it acts on states initially described in B's perspective, it is clearer to reexpress all operators relative to B. 
Recall that, for any perspective neutral operator $\hat{O}$, its B-perspective is given by $\hat{O}_{|B} = \hat{\mathcal{T}}_{B} \hat{O} \hat{\mathcal{T}}^{\dagger}_{B}$. Since $ \hat{\mathcal{T}}_{B}$ is unitary operator, we have
\begin{equation}
 \hat{\mathcal{T}}_{B} e^{i \hat{O}} \hat{\mathcal{T}}^{\dagger}_{B} =   e^{i \hat{\mathcal{T}}_{B} \hat{O} \hat{\mathcal{T}}^{\dagger}_{B}} = e^{i  \hat{O}_{|B}} .
\end{equation}
Thus each exponential factor in the $\hat{S}_{B\rightarrow A} $ can be rewritten in terms of B-perspective operator. We denote $\hat{\mathcal{T}}_{A|B}:= \hat{\mathcal{T}}_{B} \hat{\mathcal{T}}_{A}  \hat{\mathcal{T}}^{\dagger}_{B}$ as A-trivilization map expressed in terms of operators in B-perspective.  We introduce the isomorphism $\hat{\mathcal{I}}_{BA}$  which only relabels the Hilbert space, to keep track of the change of perspective:
\begin{equation}
\hat{\mathcal{I}}_{BA}: \mathcal{H}_{...|B} \rightarrow \mathcal{H}_{...|A}
\end{equation}
Therefore, we can rewrite
\begin{equation}
\hat{S}_{B\rightarrow A}  =  \hat{\mathcal{T}}_{A}  \cdot  \hat{\mathcal{T}}^{\dagger}_{B} = \hat{\mathcal{T}}^{\dagger}_{B} \cdot \left(\hat{\mathcal{T}}_{B} \hat{\mathcal{T}}_{A}  \hat{\mathcal{T}}^{\dagger}_{B}\right) = \hat{\mathcal{I}}_{BA} \cdot  \hat{\mathcal{T}}^{\dagger}_{B|B} \cdot \hat{\mathcal{T}}_{A|B}.
\end{equation}
The explicit calculation gives us
\begin{equation}\label{passiveQCFfromB}
\begin{split}
\hat{S}_{B\rightarrow A} 
& = \hat{\mathcal{I}}_{BA} \cdot e^{  \frac{i}{\hbar} \hat{\mathcal{T}}_{B}  \int d^3s   \hat{\zeta}^B_{(\mu)}(s) \left(\hat{T}^{0\mu}_A(s)+\hat{T}^{0\mu}_S(s) +2\partial_{\nu} \hat{\pi}_G^{\nu \mu} (s) + \delta_0^\mu  \frac{1}{\kappa}\Delta \hat{h}^T(s)\right) \cdot \hat{\mathcal{T}}^{\dagger}_{B}} \cdot  e^{ - \frac{i}{\hbar} \hat{\mathcal{T}}_{B} \cdot  \int d^3s   \hat{\zeta}^A_{(\mu)}(s) \left(\hat{T}^{0\mu}_B(s)+\hat{T}^{0\mu}_S(s) +2\partial_{\nu} \hat{\pi}_G^{\nu \mu} (s) + \delta_0^\mu  \frac{1}{\kappa}\Delta \hat{h}^T(s)\right) \cdot \hat{\mathcal{T}}^{\dagger}_{B}} \\
& =\hat{\mathcal{I}}_{BA} \cdot e^{  \frac{i}{\hbar}   \int d^3s   \hat{\zeta}^{B|B}_{(\mu)}(s) \left(\hat{T}^{0\mu}_{A|B}(s)+\hat{T}^{0\mu}_{S|B}(s) +2\partial_{\nu} \hat{\pi}_{G|B}^{\nu \mu} (s) + \delta_0^\mu  \frac{1}{\kappa}\Delta \hat{h}_{B|B}^T(s)\right) } \cdot  e^{ - \frac{i}{\hbar}  \int d^3s   \left( \hat{\zeta}^{A|B}_{(\mu)} (s)+\hat{\zeta}^{B|B}_{(\mu)} (s) \right) \left(\hat{T}^{0\mu}_{B|B}(s)- \hat{T}^{0\mu}_{A|B} (s)\right) } \\
&=\hat{\mathcal{I}}_{BA} \cdot  e^{ - \frac{i}{\hbar}   \int d^3s  \hat{\zeta}^{A|B}_{(\mu)} (s) \hat{T}^{0\mu}_{B|B}(s)} \cdot e^{  \frac{i}{\hbar}   \int d^3s  \hat{\zeta}^{B|B}_{(\mu)} (s) \hat{T}^{0\mu}_{A|B}(s)}  \cdot e^{ - \frac{i}{\hbar}   \int d^3s   \hat{\zeta}^{A|B}_{(\mu)}(s) \left(\hat{T}^{0\mu}_{S|B}(s) +2\partial_{\nu} \hat{\pi}_{G|B}^{\nu \mu} (s) + \delta_0^\mu  \frac{1}{\kappa}\Delta \hat{h}_{B|B}^T(s)\right) }. 
\end{split}
\end{equation}
This is \eqref{perstran} in the main text.
More generally, one may insert any unitary operator at the PN level composed by Dirac observables, to change the full system to a physically different configuration.
The \QCF{} transformation from $B$ to $A$ can be written generally as: 
\begin{equation}
    \hat{S}^{V}_{B\rightarrow A} = \hat{\mathcal{T}}_A \hat{V} \hat{\mathcal{T}}_B^\dagger,
\end{equation}
When $\hat{V} = \hat{\mathbb{1}}$, it reduces to the above Eq.(\ref{passiveQCF}). If $\hat{V}$ is chosen to be the evolution operator given by the full Hamiltonian of the system, $\hat{S}^{V}_{B\rightarrow A}$ describes the time evolution  described through a change of perspective from B to A.

To derive the transformation of the reduced observables, there are two equivalent ways to perform the calculation. The first is to map the observables in the  B-perspective back to PN description using $\hat{\mathcal{T}}^{\dagger}_B$,  and then reduced to A-perspective $\hat{\mathcal{T}}_A$. The second way is using Eq.\eqref{passiveQCFfromB} to transform all relevant quantities directly from B-perspective, and relabel the Hilbert spaces using the isomorphism map at the end. 

We illustrate the calculation for the transformation of scalar field observables $\hat{\psi}_{S|B}(\hat{X}^{(\mu)}_{B|B})$ using both approaches. 

In the first approach, through Eq.\eqref{passiveQCF}, the reduced observable of scalar field in B-perspective  transforms as : 
\begin{equation}
\begin{split}
\hat{S}_{B\rightarrow A} \cdot \hat{\psi}_{S|B}(\hat{X}^{(\mu)}_{B|B}) \cdot  (\hat{S}_{B\rightarrow A})^{\dagger} \equiv &\  \hat{\mathcal{T}}_A  \cdot \hat{\mathcal{T}}_B^{\dagger} \cdot \hat{\psi}_{S|B}(\hat{X}^{(\mu)}_{B|B}) \cdot \hat{\mathcal{T}}_B \cdot \hat{\mathcal{T}}_A^{\dagger} \\
  = &\ \hat{\mathcal{T}}_A  \cdot \hat{\psi}_{S}\left(\chi_B(\hat{\mathrm{Y}}^B_s)\right)  \cdot \hat{\mathcal{T}}_A^{\dagger} \\
    = &\ \hat{\mathcal{T}}_A  \cdot \hat{\psi}_{S}\left(\chi_A(\hat{\mathrm{Y}}^A_s)\right)  \cdot \hat{\mathcal{T}}_A^{\dagger} \\
    = &\ \hat{\psi}_{S|A}\left(s^{\mu} \hat{\mathbb{1}}_A +\hat{\zeta}^{(\mu)}_{A}(s) \right)\\
  = &\ \hat{\psi}_{S|A} \left( \hat{X}^{(\mu)}_{A|A} \right).\\
 \end{split}
 \end{equation}
In the second approach, applying Eq.\eqref{passiveQCFfromB}, we have
\begin{equation}
\begin{split}
&\hat{S}_{B\rightarrow A} \cdot \hat{\psi}_{S|B}(\hat{X}^{(\mu)}_{B|B}) \cdot  (\hat{S}_{B\rightarrow A})^{\dagger} \\
= &\  \hat{\mathcal{I}}_{BA} \cdot  e^{ - \frac{i}{\hbar}   \int d^3s  \hat{\zeta}^{A|B}_{(\mu)} (s) \hat{T}^{0\mu}_{B|B}(s)} \cdot e^{  \frac{i}{\hbar}   \int d^3s  \hat{\zeta}^{B|B}_{(\mu)} (s) \hat{T}^{0\mu}_{A|B}(s)}  \cdot e^{ - \frac{i}{\hbar}   \int d^3s   \hat{\zeta}^{A|B}_{(\mu)}(s) \hat{T}^{0\mu}_{S|B}(s)} \cdot \hat{\psi}_{S|B}(\hat{X}^{(\mu)}_{B|B}) \cdot (\hat{S}_{B\rightarrow A})^{\dagger} \\
  = & \  \hat{\mathcal{I}}_{BA} \cdot  \hat{\psi}_{S|B}\left(\hat{X}^{(\mu)}_{B|B} + \hat{\zeta}_{A|B}^{(\mu)}- \hat{\zeta}_{B|B}^{(\mu)} \right)  \cdot \hat{\mathcal{I}}^{\dagger}_{BA} \\
    = &\ \hat{\psi}_{S|A} \left( \hat{X}^{(\mu)}_{A|A} \right)
 \end{split}
 \end{equation}
For the \QCF{} A in the perspective of B,  when transform back to the perspective of A, we obtain 
\begin{equation}
\hat{S}_{B\rightarrow A} \hat{X}_{A|B}(\hat{X}_{B|B}) (\hat{S}_{B\rightarrow A})^{\dagger} 
= \ \hat{X}_{A|A} (s^{\mu} \hat{\mathbb{1}} ),
\end{equation}
as desired. We can denote $\hat{X}_{A|A} (s^{\mu} \hat{\mathbb{1}} )$ as $\hat{X}_{A|A} (s)$ for brevity.
Conversely, taking reference field B  $ \hat{X}_{B|B}(s)$  transform to the perspective of A. we have
\begin{equation}
\begin{split}
\hat{S}_{B\rightarrow A} \hat{X}_{B|B}(s) (\hat{S}_{B\rightarrow A})^{\dagger}
= &\ \hat{\mathcal{I}}_{BA} \cdot e^{-\frac{i}{\hbar} \int d^3s  \hat{\zeta}^{A|B}_{(\mu)}(\vec{s}) \hat{T}^{0\mu}_{B|B}(\vec{s})  } \cdot \hat{X}_{B|B}(s ) \cdot  e^{\frac{i}{\hbar} \int d^3s  \hat{\zeta}^{A|B}_{(\mu)}(\vec{s}) \hat{T}^{0\mu}_{B|B}(\vec{s})  } \cdot \hat{\mathcal{I}}_{BA}^{\dagger}  \\
= &\ \hat{\mathcal{I}}_{BA} \cdot \hat{X}_{B|B}(\hat{X}_{A|B} ) \cdot \hat{\mathcal{I}}_{BA}^{\dagger}  \\
= &\ \hat{X}_{B|A}(\hat{X}_{A|A}) 
\end{split}
\end{equation}
One can further expand the above equation as $\hat{S}_{B\rightarrow A} \hat{X}_{B|B}(s) (\hat{S}_{B\rightarrow A})^{\dagger} = \hat{X}_{B|A}(s) +\hat{\zeta}_{A|A}(s) + \mathcal{O}(h)$. 
In Fig.\ref{fig:zetainpersp} of the main text, we visually demonstrate the  transformation of the field $\hat{\zeta}(s)$.

Finally, let us derive the transformation on the perspectival gravitational field, yielding
\begin{equation}\label{gravityperspectival}
\begin{split}
\hat{S}_{B\rightarrow A} \cdot   \hat{g}_{\alpha \beta}^B \cdot  (\hat{S}_{B\rightarrow A})^{\dagger} 
= &\hat{S}_{B\rightarrow A} \cdot \left( \eta_{\alpha \beta}+  \hat{h}_{\alpha \beta}(\chi_B(\hat{\mathrm{Y}}^B_s)) \right)  \cdot  (\hat{S}_{B\rightarrow A})^{\dagger} \\
=  & \hat{\mathbb{1}} \eta_{\alpha \beta}+ \hat{h}_{\alpha \beta}(\chi_A(\hat{\mathrm{Y}}^A_s)) -\partial_{\alpha}\hat{\zeta}_{(\beta)}^{B|A} (\hat{X}_{A|A})-\partial_{\beta}\hat{\zeta}_{(\alpha)}^{B|A} (\hat{X}_{A|A}) \\ 
= & \hat{g}_{\alpha \beta}^A -\partial_{\alpha}\hat{\zeta}_{(\beta)}^{B|A} (\hat{X}_{A|A})-\partial_{\beta}\hat{\zeta}_{(\alpha)}^{B|A} (\hat{X}_{A|A}). 
\end{split}
\end{equation}
In order to obtain the above result, we have used the transformation on the gravitational field as
\begin{equation}
\begin{split}
&e^{ -\frac{2i}{\hbar} \int d\vec{s} \ \hat{\zeta}^{\nu}_{A|B}(\vec{s} )   \partial^{\mu} \hat{\pi}^{G|B}_{\mu \nu}(\vec{s} )  }   \cdot   \hat{h}_{\alpha \beta}(\chi_B(\hat{\mathrm{Y}}^B_s)) \cdot e^{ \frac{2i}{\hbar} \int d\vec{s} \ \hat{\zeta}^{\nu}_{A|B}(\vec{s} )   \partial^{\mu} \hat{\pi}^{G|B}_{\mu \nu}(\vec{s} )  }  \\
= &\ \hat{h}_{\alpha \beta}(\chi_B(\hat{\mathrm{Y}}^B_s)) + \frac{2i}{\hbar}  \int d\vec{s} \   [\hat{\pi}^G_{\mu \nu}(\vec{s} ) ,  \hat{h}_{\alpha \beta} (\chi_B(\hat{\mathrm{Y}}^B_s))]  \partial^{(\mu}\hat{\zeta}^{(\nu))}_{A|B}(\vec{s} )  \\
= & \ \hat{h}_{\alpha \beta}(\chi_B(\hat{\mathrm{Y}}^B_s))+ \partial_{\alpha}\hat{\zeta}_{(\beta)}^{A|B} (\chi_B(\hat{\mathrm{Y}}^B_s))+\partial_{\beta}\hat{\zeta}_{(\alpha)}^{A|B} (\chi_B(\hat{\mathrm{Y}}^B_s)). \\
\end{split}
\end{equation}
Then we need to further transform the linear shift term $\partial_{\beta}\hat{\zeta}_{(\alpha)}^{A|B} (\chi_B(\hat{\mathrm{Y}}^B_s))$ into A perspective: 
\begin{equation}
\begin{split}
& \hat{\mathcal{I}}_{BA} \cdot  e^{ - \frac{i}{\hbar}   \int d^3s  \hat{\zeta}^{A|B}_{(\mu)} (s) \hat{T}^{0\mu}_{B|B}(s)} \cdot e^{  \frac{i}{\hbar}   \int d^3s  \hat{\zeta}^{B|B}_{(\mu)} (s) \hat{T}^{0\mu}_{A|B}(s)}  \cdot \partial_{\beta}\hat{\zeta}_{(\alpha)}^{A|B} (\chi_B(\hat{\mathrm{Y}}^B_s))  \cdot  e^{ - \frac{i}{\hbar}   \int d^3s  \hat{\zeta}^{B|B}_{(\mu)} (s) \hat{T}^{0\mu}_{A|B}(s)} \cdot  e^{  \frac{i}{\hbar}   \int d^3s  \hat{\zeta}^{A|B}_{(\mu)} (s) \hat{T}^{0\mu}_{B|B}(s)} \cdot \hat{\mathcal{I}}^{\dagger}_{BA}\\
= & -  \partial_{\beta}\hat{\zeta}_{(\alpha)}^{B|A} (\chi_A(\hat{\mathrm{Y}}^A_s))  = -  \partial_{\beta}\hat{\zeta}_{(\alpha)}^{B|A} (\hat{X}_{A|A}) + \mathcal{O}(h^2), 
\end{split}
\end{equation}
which together gives the result in Eq.\eqref{gravityperspectival} and Eq.\eqref{hpitrans} in the main text.
\section{The Transformations on the Physical States}
\label{app:States}

In this section, we demonstrate how the constraint-trivialisation operator acts in an  example of a perspective-neutral state: the quantum state of gravitational field generated by a quasi-static quantum source together with the coordinate fields \cite{Chen:2024xvm}.  We will use  $\alpha = A,B,S$ labels the quantum source S and the reference fields A,B - they play an equal role regarding generating gravitational field.
\begin{equation}
 \ket{\Psi}_{ABSG} = \eta  \int \prod_\alpha \mathcal{D}(E_{\alpha})\mathcal{D}[h_{ij}] \psi (E_{\alpha}) \delta[h^T-\mathsf{h}^T_{\sum_\alpha E_{\alpha}} ] \cdot \Psi_{vac}[h_{ij}]\ket{E}_{\alpha} \ket{h_{ij}}_G
\end{equation}
in which
$\mathrm{h}^T_{\small{\sum}_{\alpha} E_{\alpha}}(\vec{x})$ is the solution of the classical  Poisson equation with the source and \QCF{}s being in the eigenstate $\ket{E}_{\alpha}$ with local eigenvalue function $E_{\alpha} (\vec x)$:
\begin{equation}\label{deth}
    \mathrm{h}^T_{\sum_{\alpha} E_{\alpha}}(\vec{x}) = \frac{\kappa}{4 \pi} \sum_{\alpha}\int d^3 y \frac{E_{\alpha}(\vec y)}{|\vec x - \vec y|}.
\end{equation}
If we only consider static source, the eigenvalues of $\hat{T}^{0i}_S$ and $\hat{T}^{0i}_B$ are negligible when acting on the state.    Since this state was obtained in temporal gauge, only the spatial part of the transformation is relevant here. Applying the constraint-trivialisation map to the static state yields
\begin{equation}\begin{split}
\hat{\mathcal{T}}_{A} \cdot  \ket{\Psi}_{ABSG} &= \eta  \int \prod_\alpha \mathcal{D}(E_{\alpha})\mathcal{D}[h_{ij}] \psi (E_{\alpha}) \delta[h^T-\mathsf{h}^T_{\sum_\alpha E_{\alpha}} ]  \Psi_{vac}[h_{ij}] \hat{\mathcal{T}}_{A} \ket{E}_{\alpha} \ket{h_{ij}}_G\\
&= \eta  \int \prod_\alpha \mathcal{D}(E_{\alpha})\mathcal{D}[h_{ij}] \psi (E_{\alpha}) \delta[h^T-\mathsf{h}^T_{\sum_\alpha E_{\alpha}} ]  \Psi_{vac}[h_{ij}] e^{-  \frac{i}{\hbar}  \int d^3s   \hat{\zeta}^A_{(0)}(\vec{s}) \left(\hat{T}^{00}_B(\vec{s})+\hat{T}^{00}_S(\vec{s}) +2\partial_{\nu} \hat{\pi}_G^{\nu 0} (\vec{s}) +   \frac{1}{\kappa}\Delta \hat{h}^T(\vec{s})\right)}  \ket{E}_{\alpha} \ket{h_{ij}}_G\\
&= \eta \int \prod_{\alpha\neq A} \mathcal{D}(E_{\alpha})\mathcal{D}[h_{ij}] \psi (E_{\alpha}) \delta[h^T-\mathsf{h}^T_{\sum_\alpha E_{\alpha}} ]  \Psi_{vac}[h_{ij}] \ket{E_A + E_B + E_S + \frac{1}{\kappa}\Delta h^T}_{A|A} \ket{E}_{\alpha | A} \ket{h_{ij}}_{G|A} \\
&= \ket{0}_{A|A} \otimes \eta'   \int \prod_{\alpha\neq A} \mathcal{D}(E_{\alpha})\mathcal{D}[h_{ij}] \tilde{\psi} (h_{ij}, E_{\alpha\neq A})  \Psi_{vac}[h_{ij}] \ket{E}_{\alpha | A} \ket{h_{ij}}_{G|A}\\
\end{split}
\end{equation}
in which $\tilde{\psi} (h_{ij}, E_{\alpha\neq A}) := \psi (-\sum_{\alpha \neq A}E_{\alpha} - \Delta h^T/\kappa, E_{\alpha\neq A}).$
In the last line, we use the functional delta to perform the $E_A$ integration. 
The corresponding functional Jacobian is field-independent and abosrbed into the normalisation $\eta'$. The physical state in the perspective of A is expressed as 
\begin{equation}
 \ket{\Psi}_{BSG|A} =  \eta'   \int \prod_{\alpha\neq A} \mathcal{D}(E_{\alpha})\mathcal{D}[h_{ij}] \tilde{\psi} (h_{ij}, E_{\alpha\neq A})  \Psi_{vac}[h_{ij}] \ket{E}_{\alpha | A} \ket{h_{ij}}_{G|A}
\end{equation}
After the constraint trivialisation map, the state of the reference field factorized out, as expected:
\begin{equation}
\hat{\mathcal{T}}_{A}|\Psi\rangle_{ABSG}  = |E_A =0 \rangle_{A|A}  \otimes |\Psi\rangle_{BSG|A}.
\end{equation}
In general, if $\hat{T}^{0i}_A$  is not negligible as a quantum source for the quantum state, we will have
\begin{equation}
\hat{\mathcal{T}}_{A}|\Psi\rangle_{ABSG}  = |T^{0\mu}_A  =0 \rangle_{A|A}  \otimes |\Psi\rangle_{BSG|A}.
\end{equation}
Here $T^{0\mu}_A = \sum_{\nu} T^{0\mu}_{(\nu)}$.
Under the conditions of Eq. (\ref{T0icondi}, \ref{T00condi}), $|T^{0\mu}_A \rangle := \otimes_{\mu}|T^{0\mu}_A\rangle \approx |E \rangle_A \otimes_i |\Pi^{i} \rangle_A$, which means that the eigenbasis $|T^{0\mu}_\alpha \rangle$ can be well-defined  when we keep the leading order of the stressenergy tensor under condition Eq.(\ref{T0icondi}, \ref{T00condi}). In general, one cannot diagonalize  $T^{0\mu}_A$ on an equal-time hypersurface due to the Schwinger terms. 

Changing from A’s perspective to B’s, we obtain
\begin{equation}
\begin{split}
\hat{S}_{A\rightarrow B} |0\rangle_A  \otimes |\Psi\rangle_{BSG|A} &=  \hat{\mathcal{T}}_{B} \cdot \hat{\mathcal{T}}^{\dagger}_{A} |0\rangle_A  \otimes |\Psi\rangle_{BSG|A}  = \hat{\mathcal{T}}_{B} |\Psi\rangle_{ABSG} \\
& = \eta'   \int  \mathcal{D}[E_{A|B}] \mathcal{D}[E_{S|B}]\mathcal{D}[h^B_{ij}] \tilde{\psi} (h^B_{ij}, E_{S|B}, E_{A|B})  \Psi_{vac}[h_{ij}] \ket{E}_{S| B} \ket{E}_{A| B}  \ket{h_{ij}}_{G|B}
\end{split}
\end{equation}

\section{Von Neumann Measurement Scheme Relative to a  Quantum Reference Field}
\label{app:vNeum}

\subsection{Brief introduction to Von Neumann measurement scheme }
\label{app:vNeumIntro}
In quantum mechanics, the von Neumann (vN) measurement scheme provides a way to understand the measurement procedure by coupling a measurement device, say $M$, to the system $S$ one wants to measure. The setup is the following:
\begin{enumerate}
    \item Long before the measurement starts, the quantum state of $M$ and $S$ are prepared in a product state, i.e.\,$\hat{\omega}_0^M \otimes \hat{\rho}_S$, where $\hat{\omega}_0^M$ is a fiducial state that is chosen conveniently depending on the physical situation.
    \item At some time $t_0$, an impulsive interaction couples $M$ and $S$. This can be written using a Hamiltonian formulation as $\hat{H}_{int}= \hat{O}_{MS}\delta(t-t_0)$, where $\hat{O}_{MS}$ is the joint observable on $M$ and $S$. Notice that the action of $\hat{H}_{int}$ can be written on the state as
    \begin{equation}
        \begin{split}
            \mathcal{C}_{MS}\left(\hat{\omega}_0^M \otimes \hat{\rho}_S\right) &= \overleftarrow{\mathrm{T}}\left( e^{-\frac{i}{\hbar}\int dt \hat{H}_{int} (t)}\right)\left(\hat{\omega}_0^M \otimes \hat{\rho}_S\right)\overrightarrow{\mathrm{T}} \left(e^{\frac{i}{\hbar}\int dt \hat{H}_{int}(t)}\right)=\\
            &= e^{-\frac{i}{\hbar}\hat{O}_{MS}}\left(\hat{\omega}_0^M \otimes \hat{\rho}_S\right)e^{\frac{i}{\hbar}\hat{O}_{MS}},
        \end{split}
    \end{equation}
    where $\overleftarrow{\mathrm{T}}(\cdot)$ and $\overrightarrow{\mathrm{T}}(\cdot)$ indicate the time-ordering operator, with the arrows pointing from the earlier to the later times.
    \item Finally, one obtains the probability of detecting an outcome $n$ by projecting on a state $\ket{\phi_n}_M$ of the device. Hence, overall we have
    \begin{equation}
        p_n = \mathrm{Tr}\left[\mathcal{C}_{MS}\left(\hat{\omega}_0^M \otimes \hat{\rho}_S\right)\hat{\Pi}_n^M \right],
    \end{equation}
    with $\hat{\Pi}_n^M = \ket{\phi_n}_M \bra{\phi_n}$.
\end{enumerate}
For example, if $\hat{O}_{MS}= \hat{p}_M \otimes \hat{O}_S$, with $\hat{p}_M$ being the momentum of the device, we choose as our fiducial state the perfectly localized state at position zero, i.e.\,$\hat{\omega}_0^M = \ket{x=0}_M \bra{x=0}$. If the system is not in an eigenstate of $\hat{O}_S$, the measurement in general entangles the system and the device. This can be easily seen by choosing a basis $\ket{\lambda_n}_S$ of eigenvectors of $\hat{O}_S$, i.e.\,$\hat{O}_S \ket{\lambda_n}_S = \lambda_n \ket{\lambda_n}_S$. By expanding the state of $S$ in that basis (we take here a pure state for brevity), we obtain 
\begin{equation}
    e^{-\frac{i}{\hbar}\hat{p}_{M}\otimes \hat{O}_S}\ket{x=0}_M \sum_n c_n \ket{\lambda_n}_S = \sum_n c_n \ket{\lambda_n}_M \ket{\lambda_n}_S,
\end{equation}
where $c_n$ are arbitrary coefficients. Finally, the projection over some state of $M$, e.g. $\ket{x=\lambda_m}_M$, gives the probability of finding $S$ in the eigenvector $\ket{\lambda_m}_S$, i.e. $p_m = |c_m|^2$.

\subsection{The measurement scheme in a  quantum reference field}

Let us now explain  how the von Neumann measurement scheme in Appendix \ref{app:vNeumIntro} extends to the case of \QCF{}s.  In general, measurements in quantum field theory are very subtle.  For instance, it has been shown that superluminal signalling and lack of covariance arise from a too simplistic use of instantaneous state-update rules~\cite{Sorkin1993, HellwigKraus1969, HellwigKraus1970a, HellwigKraus1970b}. A coherent framework for measurements in relativistic settings remains an active area of research~\cite{FewsterVerch2020, Papageorgiou:2023nvf}.  Significant progress has been made in a series of recent works~\cite{MartinMartinez2015, FewsterVerch2020, Bostelmann:2020unl, deRamonPapageorgiouMartinMartinez2023, Gisin:2023dck,Fewster:2024pur, Simmons:2025cmv}, to which we refer the readers for the current state-of-the-art discussion on these topics.
Devising a complete relativistic relational measurement theory would require a dedicated (series of) works on its own, and hence we leave it to future work. Our goal here is to show that the reduced observables constructed in the main text can, in principle, be accessed in a von Neumann-type measurement scheme using a relational construction.

We restrict attention to a single \QCF{} $A$ and a system $S$, leaving gravity and the other \QCF{} $B$ implicit. As in the Page--Wootters framework~\cite{Mondragon:2007xn, Giovannetti:2015qha, castro2020quantum, Hausmann2025measurementevents}, a measurement device  is needed to record the measurement outcome and keep the post-measurement state within the constraint surface. Unlike in the Page--Wootters setting, however, the device is treated as a physical system with its own dynamics and enters the gravity constraints. 
Specifically, we introduce a probe $\hat{\phi}_P$, which plays the role of measurement device, described by
a physical scalar field, with its canonically conjugated momenta $\hat{\pi}_P$. We also introduce a quantum clock  degree of freedom $\hat{\tau}_M$
which triggers the measurement interaction. 
  The probe and clock  enter the constraints in the same way as the system. 
  
  The total  Hamiltonian $\hat{H}_{tot}$ of Eq.\,\eqref{eq:FullHam}  in the main text acquires an additional interaction term, 
A suitable relational interaction Hamiltonian is
\begin{equation}
    \hat{H}_{int} = \int d^3\vec{s} \,\delta \left(\hat{\tau}_{C}(\chi_A(\hat{\mathrm{Y}}^{(0)}_s))-t^* \hat{\mathbb{1}}\right)  \,\hat{\psi}_{S}(\chi_A(\hat{\mathrm{Y}}_s)) \hat{\phi}_{P}(\chi_A(\hat{\mathrm{Y}}_s)).
\end{equation}
in which $\hat{\psi}_{S}(\chi_A(\hat{\mathrm{Y}}_s))$, $\hat{\phi}_{M}(\chi_A(\hat{\mathrm{Y}}_s))$ and $\hat{\tau}_{C}(\chi_A(\hat{\mathrm{Y}}^{(0)}_s))$ are diffeomorphism-invariant relational operators respectively between the system, the probe, and the clock and the \QCF{} A. 
For simplicity, and to illustrate the structure of the relational measurement coupling, we omit the smearing functions in the expression.

We assume that the clock $\hat{\tau}_{C}$ is a good clock in the region where the measurement takes place. In particular, its eigenvalue configurations should be smooth and locally monotonic along the $s^0$ direction.
For simplicity, 
we further assume that the temporal \QCF{} $X^{(0)}_A$ has constant eigenvalue on each $\Sigma_{s^0}$ hyper-surface. 
Physically, the the interaction switches on when the clock shows time $t^*$. Formally, this is achieved via the delta function which, when applied to the state of the clock and the \QCF{}, is different to zero when the eigenvalue of the relational operator is $\hat{\tau}_{C}(\chi_A(\hat{\mathrm{Y}}^{(0)}_s))$ is $t^*$.
 The background label $s^0$ only parametrizes the Heisenberg picture operator; it is not the physical time at which the measurement occurs. 
The interaction  may be supported on different $s^0$-hypersurfaces in different clock superposition branches, but at the same time $t^*$ according to the clock. This is a similar mechanism to Refs.\,\cite{Zych:2017tau, castro2020quantum, giacomini2021spacetime}.

Now we switch to the perspective of \QCF{} A. Using the reduction of relational observables,
the action of the trivialisation map on the interaction Hamiltonian yields 
\begin{equation}\label{HintA}
    \hat{H}_{int|A}(s^0)=\hat{\mathcal{T}}_A\hat{H}_{int}\hat{\mathcal{T}}_A^\dagger = \int d^3\vec{s}\, \delta\left(\hat{\tau}_{C}(\hat{X}^{(0)}_{A|A}(s))-t^* \hat{\mathbb{1}}\right) \hat{\psi}_{S|A} (\hat{X}_{A|A}(s))  \hat{\phi}_{P|A}(\hat{X}_{A|A}(s)),
\end{equation} 
i.e.\,we obtain the reduced Hamiltonian composed by the relational observables reduced to the \QCF{} $A$.Under the same trivialisation map, the physical state factorizes as
(see Appendix~\ref{app:Tri})
\begin{equation}
    \hat{\mathcal{T}}_A \ket{\Psi}_{ph} = \ket{T^{0\mu}_A=0}_{A|A} \ket{\psi}_{\text{rest}|A}.
\end{equation} 
Here ``rest” denotes all non-A degrees of freedom in the full system.
We now prepare the initial states of the system $\ket{\Psi_0}_{S|A}$ , clock $\ket{\varphi}_{C|A}$, and the probe $\ket{\sigma}_{P|A}$ as a product state in the perspective of A:
\begin{equation}
\ket{\Phi_0}_{|A}=|\Psi_0\rangle_{S|A}
\otimes
|\sigma\rangle_{P|A}
\otimes
|\varphi\rangle_{C|A}.
\end{equation}
The probe is prepared in an eigenstate of the canonically conjugated momentum to $\hat{\phi}_{P|A}$:$\ket{\sigma}_{P|A}=\ket{\pi}_{P|A}$.
Strictly speaking, momentum eigenstates of the measurement device are not normalized, but in realistic situations such states approximate sharply peaked wave packets in momentum basis, which are instead correctly normalized. 
We work in the interaction picture with respect to the free clock evolution. The general quantum clock state $|\varphi\rangle_C$ is expanded in a time-independent branch basis:  $|\varphi\rangle_C = \int d\tau\varphi_C(\tau) |\tau\rangle_C$.  Here $\tau$ labels a clock eigenstate; 
In the Heisenberg picture, the clock operator is diagonal in this basis, written in the perspective of \QCF{} A as:
\begin{equation}
\hat{\tau}_{C|A} (x^0_A) |\tau\rangle_{C|A} =\tau (x^0_A) |\tau\rangle_{C|A} 
\end{equation}
Since we have required that the eigenvalue of  $\hat{\tau}_C$ is monotonous in the temporal direction, every eigenfunction $\tau (x^0_A)$ is an invertible function of $x^{(0)}_{A}$ or $s^{(0)}$. Therefore $x^{(0)}_{A} = \tau^{-1}(t^*)$ is the corresponding time in terms of the quantum reference field $\hat{X}^{(0)}_A$.  The interaction between the system and the device is triggered at the clock reading of $t^*$, which corresponds to a superposition of reference field value $x^{(0)}_A$.

The full state $\ket{\Phi_t}_{|A}$ at time $t$ in the interaction picture is
\begin{equation}\ket{\Phi_t}_{|A}=\overrightarrow{\mathrm{T}}\left( e^{-\frac{i}{\hbar}\int_0^t ds^0 \hat{H}_{int|A} (s^0)}\right) \ket{T_{0\mu}^A=0}_{A|A} \ket{\Psi_0}_{S|A}\ket{\sigma}_{P|A} \ket{\varphi}_{C|A},
\end{equation}
We then remove the remaining redundancy of the description by conditioning on a specific eigenstate of the \QCF{} $\ket{x}_{A|A}$.
All the operator-valued operators, for instance, $\hat{\psi}_{S|A} (\hat{X}_{A|A})$, become their gauge-fixed version $\hat{\psi}_{S|A} (x_{A})$.  The interacting Hamiltonian reduced to the $A$ perspective in Eq.\,\eqref{HintA} becomes
\begin{equation}
\hat{H}_{int|A} (x^0_A)=
\int \mathcal{D}[\tau] \int d \mu( \vec{x}_A)\, \delta\left(\tau_{C|A}(x^0_A)-t^* \hat{\mathbb{1}}\right) \hat{\psi}_{S|A} (x^0_A,\vec{x}_A) \hat{\phi}_{P|A}(x^0_A,\vec{x}_A) \otimes |\tau \rangle_{C|A} \langle \tau|,
\end{equation}
where we have changed variables in the integral from $\vec{s}$ to $\vec{x}_A$. Notice that the  Jacobian of the transformation is absorbed into the integration measure $\mu( \vec{x}_A)$. The crucial point is that for each clock eigenstate $|\tau '\rangle$,
the delta function selects the point $x^0_A$ at which that clock configuration reads $t^*$. Therefore the final state remains a superposition over clock configurations, with each clock amplitude producing a different relational interaction time $x_A^0= \tau^{-1}(t^*)$.
After the interaction, we obtain schematically
\begin{equation}
\begin{split}
    \ket{\Phi_t}_{|A}&= \overrightarrow{\mathrm{T}}\left( e^{-\frac{i}{\hbar}\int_0^t dx_A^{0} \hat{H}_{int|A} (x_A^{0})}\right)  \ket{\Psi_0}_{S|A}\ket{\pi}_{P|A}  \int \mathcal{D}[\tau] \  \varphi_{C|A}(\tau (x^0_A)) |\tau \rangle_{C|A}\\
    &= \eta  \int  \mathcal{D}[\tau]   e^{-\frac{i}{\hbar}\int d \mu( \vec{x}_A) J_{\tau} \hat{\psi}_{S|A} (\tau^{-1}(t^*), \vec{x}_A)  \hat{\phi}_{P|A}(\tau^{-1}(t^*), \vec{x}_A) } \varphi_{C|A}(\tau)\ket{\Psi_0}_{S|A}\ket{\pi}_{M|A} \ket{\tau}_{C|A}.
    \end{split}
\end{equation}
in which $J_{\tau}$ is the  Jacobian $J^{-1}_{\tau}:= |\partial_0\tau(x_A^{(0)})|_{x_A^{(0)} = \tau^{-1}(t^*)}$ from the integration of the delta function. In what follows, we simply absorb it into the integration measure,  
since it is not relevant to derive the final results.  

We now expand the system $\ket{\Psi_0}_{S|A}$ in its field basis, i.e.\,$\ket{\Psi_0}_{S|A} = \int  \mathcal{D}[\psi] \Psi_0[\psi]\ket{\psi}_{S|A}$, the post-interaction state is
\begin{equation}
    \ket{\Phi_t}_{|A}= \eta \int  \mathcal{D}[\tau] \int \mathcal{D}[\psi] \Psi_0[\psi] \varphi_{C|A}[\tau] \ket{\psi}_{S|A}\ket{\pi +  \psi (\tau^{-1}(t^*))}_{P|A}\ket{\tau}_{C|A}.
\end{equation}
The previous expression does not encode information about the background coordinates $s^\mu$, but the state is only defined in terms of the physical time of the clock. Depending on how we treat the clock, we can read out different information about the system. Therefore, the clock is not a relatively measurable reference in the sense defined in \cite{Vilasini:2025qun}.

If we do not condition on the clock configuration, then the measurement statistics of the system contains a clock-weighted average over the wavefunction of the clock, and hence a mixture of  different $\tau^{-1}(t^*)$.  
If we condition on the clock, namely we project on $\hat{\Pi}_{\tau}^{C|A}=  \ket{\tau_n}_{C|A} \bra{\tau_n}$. For a continuous clock-configuration basis, this projector should be understood formally, or as a coarse-grained projector. and on the quantum state of the probe on $\hat{\Pi}_n^{P|A} = \ket{\pi_n}_{P|A} \bra{\pi_n}$,
then the conditional probability is 
\begin{equation}
\begin{split}
p_{n|\tau} &= \frac{\mathrm{Tr}_{SCP}\left[\mathcal{C}_{SCP|A}\left(\hat{\omega}_{P|A} \otimes \hat{\rho}_{S|A}\otimes \hat{\varrho}_{C|A}\right)\left(\hat{\Pi}_n^{P|A} \otimes \hat{\mathbb{1}}_{S|A}\otimes \hat{\Pi}_{\tau}^{C|A}  \right) \right]}{\mathrm{Tr}_{SCP}\left[\mathcal{C}_{SCP|A}\left(\hat{\omega}_{P|A} \otimes \hat{\rho}_{S|A}\otimes \hat{\varrho}_{C|A}\right)\left(\hat{\mathbb{1}}^{P|A} \otimes \hat{\mathbb{1}}_{S|A}\otimes \hat{\Pi}_{\tau}^{C|A} \right) \right]},\\
\end{split}
\end{equation}
where $\mathcal{C}_{SCP|A}(\cdot) = \overleftarrow{\mathrm{T}}\left( e^{-\frac{i}{\hbar}\int_0^t dx_A^0 \hat{H}_{int|A} (x_A^0)}\right)(\cdot)\,\overrightarrow{\mathrm{T}}\left( e^{\frac{i}{\hbar}\int_0^t dx_A^0 \hat{H}_{int|A} (x_A^0)}\right)$ is defined in analogy to the quantum mechanical case,  $\hat{\omega}_{P|A} = \ket{\sigma}_{P|A}\bra{\sigma}$, $\hat{\varrho}_{C|A} = \ket{\varphi}_{C|A}\bra{\varphi}$, and $\hat{\rho}_{S|A}= \ket{\Psi_0}_{S|A}\bra{\Psi_0}$.

It can be checked that the conditional probability can be equivalently written as
\begin{equation}
p_{n|\tau} = |\Psi_0\left[\psi\left(\tau^{-1}(t^*), \vec{x}_A\right)\right]|^2, \text{when} \ |\pi_n\rangle_{P|A}=\ket{\pi +  \psi (\tau^{-1}(t^*))}_{P|A}
\end{equation}
In summary, we find a correspondence between this measurement scheme and its original particle version: the probability of obtaining a certain outcome gives, over many rounds of measurements, the amplitude of the wavefunctional. In addition, the eigenvalue of $\hat{\psi}_S(x_A)$ can be inferred from the shift of the pointer state of the probe P: the difference between the final and initial eigenvalue reading determines the corresponding eigenvalue of the relational field.

Overall, we find that it is possible to construct a Von Neumann measurement procedure, where the relational observables look like standard observables in the reduced frame of an arbitrary \QCF{}. In particular, this procedure is fully relational, in that it does not require, at any point, to refer to the value of the background coordinates $s$. The interaction is quantum controlled by a physical clock, and the reading of relational observables comes from measuring the probe.   
Here, we have not addressed how this measurement procedure could be seen in a different \QCF{}. Structurally, the transformation would yield something along the lines of the Von Neumann measurement scheme for QRFs introduced in Ref.\,\cite{Giacomini:2017zju}, although adapted to quantum fields. Notice, however, that every discussion about measurements in QRFs always assumes perfect communication between the QRFs, in order to avoid situations like the Wigner's friend thought experiment~\cite{wigner1995remarks, Brukner:2018oda, Frauchiger:2018zma, bong2020strong}. A full treatment of the measurement in QRFs should also take into account this setup, and constitutes a whole line of research. Understanding how the current discussions on the Wigner's friend scenario fit into our analysis is an interesting direction to be addressed in future work.

\end{widetext}

\bibliography{biblioQD}
\end{document}